\documentclass[10pt,superscriptaddress,english,showpacs,longbibliography,nofootinbib,floatfix]{revtex4-1}
\pdfoutput=1
\usepackage{amsmath}
\usepackage{amssymb}
\usepackage{dsfont}
\usepackage[super]{nth}
\usepackage{amsthm}
\usepackage{amsfonts}
\usepackage{listings}
\usepackage{longtable}
\usepackage{enumerate}
\usepackage{latexsym}
\usepackage{color}
\usepackage{multirow}
\usepackage[figuresright]{rotating}
\usepackage[table]{xcolor}
\usepackage{setspace} 
\usepackage{blindtext}
\usepackage{dcolumn}\usepackage{braket}

\usepackage{array,etoolbox}
\preto\tabular{\setcounter{magicrownumbers}{0}}
\newcounter{magicrownumbers}

\usepackage{bm}
\usepackage{hyperref}
\hypersetup{
 pdfnewwindow=true, colorlinks=true,
 linkcolor=blue, anchorcolor=blue,
 citecolor=blue, filecolor=blue,
 menucolor=blue, urlcolor=blue}
\definecolor{red}{rgb}{1,0,0}
\definecolor{blue}{rgb}{0,0,1}
\definecolor{black}{rgb}{0,0,0}

\usepackage{chemformula}

\newcommand{\Pso}{P$_{\textrm{\footnotesize{SO}}}$}
\newcommand{\Ps}{P$_{\textrm{\footnotesize{S}}}$}

\newcommand{\Pnt}{P$_{\textrm{\footnotesize{NT}}}$}
\newcommand{\Pnte}{P$_{\textrm{\footnotesize{NTE}}}$}
\newcommand{\Psoin}{P$_{\textrm{\footnotesize{SOintr}}}$}
\newcommand{\Psog}{P$_{\textrm{\footnotesize{SOG}}}$}

 \usepackage{appendix}

\begin{document}

\title{Automatic calculation of symmetry-adapted tensors under spin-group symmetry. STENSOR, a new tool of the Bilbao Crystallographic Server}

\author{Luis Elcoro}
	\email{luis.elcoro@ehu.eus}
\affiliation{Department of Physics, Faculty of Science and Technology, UPV/EHU, Bilbao, Spain}

\author{Jesus Etxebarria}
	\email{j.etxeba@ehu.eus}
\affiliation{Department of Physics, Faculty of Science and Technology, UPV/EHU, Bilbao, Spain}

\author{J. Manuel Perez-Mato}
\email{jm.perezmato@gmail.com}
\affiliation{Faculty of Science and Technology, UPV/EHU, Bilbao, Spain}

	\author{Emre S. Tasci}
\affiliation{Department of Physics Engineering, Hacettepe University, 06800 Ankara, Turkey}

\begin{abstract}
	We present STENSOR, a new computational tool integrated into the Bilbao Crystallographic Server, designed for the automatic calculation of symmetry-adapted tensors under spin group symmetry. The program requires either a file containing the structural data of the magnetic compound or the generators of the oriented spin point group, together with the so-called generalized Jahn symbol associated to the tensor of interest. The user can propose any arbitrary tensor type or select a particular one from a predefined list. The program output returns the symmetry-adapted tensor under the spin point group and also under the magnetic point group, which is also calculated. The comparison of these two tensor forms allows to distinguish the coefficients that are due to spin-orbit coupling effects from those that have a non-relativistic origin and thus are usually more important. A couple of examples are given to illustrate the operation of the program. \end{abstract}

\maketitle

\section{Introduction}
\label{sec:introduction}
Spin groups describe the symmetries of magnetic materials in the non-relativistic limit, i.e., in the absence of spin-orbit coupling (SOC) \cite{Litvin1974,Litvin1977,yuan2020,liu2022}. Although they are approximate symmetries, spin groups have attracted considerable interest in recent years \cite{Chen2024,jiang2024,xiao2024}, as they account for properties associated with physically significant effects. By contrast, effects arising from SOC are typically much weaker. Comparing the properties allowed by magnetic groups, which are exact symmetries, with those permitted by spin groups thus helps to distinguish features resulting from SOC from those that represent dominant, robust effects. For instance, when dealing with physical properties described by tensors, coefficients that originate from SOC vanish under spin-group symmetry but may remain nonzero under the corresponding magnetic group.

In this context, we have recently reviewed the transformation properties that the most important crystal tensors must satisfy under spin group symmetry \cite{etxebarria2025}. By generalizing Neumann’s principle to spin point groups (SpPGs), we have determined the symmetry-adapted forms of key tensors describing equilibrium, transport, and optical properties. We have found that tensor transformation rules under spin symmetry are significantly more varied and complex than those under magnetic symmetry.

Given a crystal whose symmetry is described by a specific spin space group (SpSG), the constraints that this 
symmetry imposes on a physical property represented by a tensor can be obtained by requiring the tensor to be invariant under the operations of the SpPG associated with the SpSG. The SpPG consists of operations of the form $\{U||R\}$, combining a spin part, represented by the matrix $U$, and a space part, represented by the matrix $R$. These matrices act on the components of a rank-$r$ tensor $A$, transforming them in ways that depend on the nature of the tensor. In any case, the transformation can always be encoded through a combination of $r$ labels V and M, together with the letters $e$ and/or $a$ in some cases \cite{etxebarria2025}, which characterize how each tensor component transforms under the action of $U$ and $R$. These combinations generalize the so-called Jahn symbols, commonly used for magnetic point groups (MPGs) \cite{jahn1949,Gallego2019}. Upon an operation $\{U||R\}$ the transformation of the tensor involves the matrix $R$ if the label is V and the matrix $U$ for the M label. The letters $e$ and $a$ indicate, respectively, a change of sign in the transformation if $R$ and $U$ are improper operations. In other words,  $e$ and  $a$  denote  that the tensor is  even under space inversion and odd under time reversal, respectively. On some occasions square or curly brackets are included in the symbol, indicating symmetry or antisymmetry of pairs of indices. In the context of spin groups two contributions to the magnetization can be distinguished \cite{Watanabe2024,etxebarria2025}, the spin contribution ${\bf M}^s$ , which is a rank-1 tensor of type M, and the orbital contribution ${\bf M}^{orb}$, which is of type $ae$V. These symbols mean that $\{U||R\}$ transforms the components of ${\bf M}^s$ and ${\bf M}^{orb}$ into $M_i^{s\prime}=U_{ij}M_j^s$ and $M_i^{orb\prime}=\det(U)\det(R) R_{ij} M_j^{orb}$, respectively ($\det$ stands for determinant). Similarly, the spin contribution to the magnetoelectric tensor (inverse effect), which connects an applied electric field ${\bf E}$ with an induced spin magnetization ${\bf M}^s$ ($M_i^s=\alpha_{ij}^T E_j$)  is of type MV, since an operation $\{U||R\}$ transforms vectors ${\bf M}^s$ and ${\bf E}$ according to matrices $U$ and $R$ respectively, and therefore, the new $\alpha^T$-tensor must have components $\alpha_{mn}^{T\prime}=U_{mi}R_{nj}\alpha_{ij}^T$. Analogously, a more complicated Jahn symbol such as $ae$MV\{V$^2$\} means that $\{U||R\}$ transforms a 4-rank tensor $A_{ijk\ell}$ of components into
\begin{equation}
	A_{mnpq}^{'}=\det(U)\det(R)U_{mi}R_{nj}R_{pk}R_{q\ell}A_{ijk\ell}
\end{equation}
and is antisymmetric in the \nth{3} and \nth{4} indices. Here, all subindices range from 1 to 3, and all the components of $A^{’}$, $A$, $R$, and $U$ are referred to the same orthonormal frame.

In ref. \citet{etxebarria2025}, approximately 40 different generalized Jahn symbols are listed to account for the most common tensor properties. Given this wide range of transformation behaviors, it seems desirable to have a tool capable of performing these calculations automatically. In this paper, we present such a tool, which we have named STENSOR, reflecting its clear parallelism with the program MTENSOR, used to obtain tensors adapted to the symmetry under magnetic point groups \cite{Gallego2019}. STENSOR is available through the Bilbao Crystallographic Server at the following address: \href{https://www.cryst.ehu.es/cryst/stensor.html}{cryst.ehu.es/cryst/stensor.html}.

A similar package for handling tensors which describe a few transport and optical processes has recently been reported in \citet{xiao2025} \section{The program}
\label{sec:program}
STENSOR operates in a similar way to MTENSOR: given a specific tensor and a SpPG, the program returns the form of the tensor adapted to the symmetry of the specified group. The input data can be introduced in two alternative ways, referred to as Option A and Option B. In Option A, a file containing the structural data of the magnetic compound must be uploaded. Two file formats are supported: scif (spin cif) and mcif (magnetic cif). In Option B, the symmetry operations of the SpPG are entered manually.

\begin{figure}[ht]
	\centering
	\includegraphics[width=0.6\textwidth]{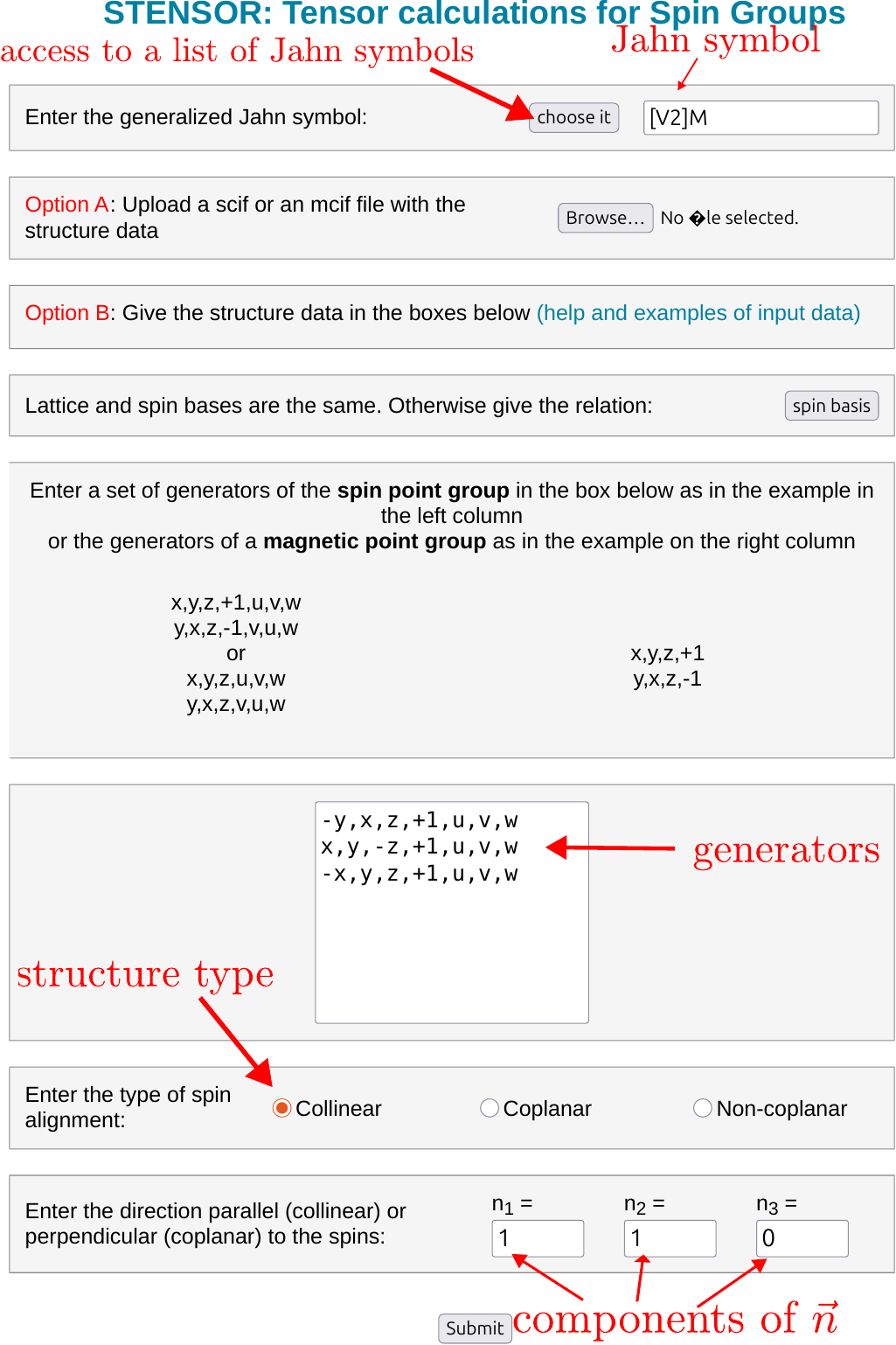}
	\caption{\label{fig:menu}Input page of STENSOR. The parameters included correspond to the example of section \ref{sec:example}.}
\end{figure}

In both cases, the Jahn symbol corresponding to the tensor of interest must also be specified (top right of Fig. \ref{fig:menu}). STENSOR provides a list of the most common properties, including the corresponding generalized Jahn symbols and the constitutive equations that define the tensors. Nevertheless, in general, the user can construct any arbitrary Jahn symbol from scratch.

If Option A is selected and a so-called \emph{scif} file is provided, the input is already complete, since this file contains all the information needed to determine the symmetry-adapted tensor forms (in particular all the symmetry operations of the SpPG are explicit). Alternatively, if an \emph{mcif} file is uploaded, STENSOR assumes a \emph{minimal} SpPG \cite{etxebarria2025}, i.e., a SpPG composed of the MPG operations together with possible spin-only operations characteristic of collinear and coplanar spin arrangements, whose presence is verified by the program. In this case as well, the input is complete. In this context, it should be noted that the format for scif files with magnetic structures described under their SpSGs is still under development, as an extension of the mcif format, where MSGs are employed. A preliminary version of the scif format is already in use by the program FINDSPINGROUP \cite{Chen2024} and is supported by the latest version of Jmol \cite{hanson2013}. Integration  of scif files into the MAGNDATA database \cite{Gallego2016} is planned for the near future.

In contrast, in Option B the SpPG operations are introduced manually by first giving a set of generators of the SpPG associated with the so-called nontrivial SpSG as 6-component vectors $x,y,z,u,v,w$ (\emph{symcard} format). The first three components refer to the lattice space $(x,y,z)$  and the last three to the spin space $(u,v,w)$. The lattice coordinates are expressed in a lattice basis $({\bf a},{\bf b},{\bf c})$ defined by the conventional unit cell and the axes normally used for the description of space operations. Thus, for example, trigonal and hexagonal lattices are always referred to an (oblique) hexagonal basis. The spin operations are in general defined in a different basis $({\bf a}_s,{\bf b}_s,{\bf c}_s)$ although, by default, they are referred to the same $({\bf a},{\bf b},{\bf c})$ basis even in trigonal or hexagonal lattices. An option, however, exists to choose an independent spin reference frame by means of a transformation matrix. This option can be useful for example if the spin operations are described in an orthonormal reference system independently of the crystallographic frame, which is the convention adopted by many authors \cite{Chen2024,jiang2024,xiao2024}. Alternatively to the 6-component vector format, the generators can be introduced in a slightly modified format of 7-component vectors $x,y,z,\pm1,u,v,w$ ,where the $\pm1$ component is the determinant of the matrix that represents the operation in the spin space, $\det(U)$. In this form, $-1$  indicates explicitly that time reversal is included in the operation. Although this additional information is redundant, this format can be useful for obtaining more efficiently the transformation laws of tensors whose Jahn symbols do not contain the letter M. These cases can be described by means of an effective MPG and do not require the explicit form of the $U$ operation, as only whether it is proper or improper is relevant.

A simpler possibility can also be used within OptionB: instead of the previously described 6 or 7-component vector format, one can input a set of generators of an MPG using the usual 4‑component vector format $(x,y,z,\pm1)$. The program then interprets that the SpPG to be considered is obtained simply by adding to the MPG the spin‑only symmetry associated with the collinearity or coplanarity of the spin arrangement. In a non‑coplanar case, the assumed SpPG and the input MPG coincide. This format is particularly useful when working with minimal SpPGs, which statistically account for about 75\% of magnetic structures \cite{Gallego2016,Chen2024,etxebarria2025}.

Once the SpPG associated with the nontrivial SpSG is known, the SpPG symmetry information must be completed by giving the type of magnetic structure (collinear, coplanar or noncoplanar), i.e., the so-called intrinsic spin-only group. This group is formed by elements of the form $\{U||1\}$, where $U$ belongs to the groups $^{\infty_{\bf n}m}1$, $^{m_{\bf n}}1$ or 1 for the collinear, coplanar or non-coplanar cases, respectively. Here $^{\infty_{\bf n}m}1$  denotes the continuous  group of all rotations around the direction of the spins together with all mirror planes containing this direction, while  $^{m_{\bf n}}1$ consists of the identity and a mirror plane with the orientation of the spin plane. In the collinear and coplanar cases the spin direction or the normal to the spin plane ${\bf n}$ must be also provided. This direction is always given in the $({\bf a}_s,{\bf b}_s,{\bf c}_s)$ basis.

The output of the program consists of the following information:
\begin{itemize}
\item Setting used to express the space operations present in the SpPG, and crystal system. If a nonstandard setting is used, the transformation matrix to the standard setting is provided. The information for deducing the crystal system is extracted from the space-part operations of the SpPG.
\item Full set of symmetry operations of the nontrivial point group, deduced from the input generators.
\item Identified MPG as a subgroup of the SpPG, and the Jahn symbol of the tensor under the MPG symmetry. The MPG is recognized from the set of SpPG operations of the form $\{\pm R||R\}$, and its Jahn symbol can be easily derived from the symbol given for the SpPG by performing the substitution M$\to ae$V \cite{etxebarria2025}.
\item Complete form of the symmetry adapted tensor both under the MPG and SpPG. Depending on the space group of the space operations of the SpPG, different choices are possible for the reference frame used to express the tensor. In all cases an orthonormal frame is employed, which is also explicitly given.
\end{itemize} \section{Examples}
\label{sec:examples}
We now present two examples of different complexity to illustrate key features of the program. The examples presented here were generated using Option B for the input data. Both examples correspond to real structures which have been taken from the MAGNDATA database \cite{Gallego2016}. \subsection{MnF$_2$ (entry 0.15 in MAGNDATA)}
\label{sec:MnF}
\begin{figure}[ht]
	\centering
	\includegraphics[width=0.8\textwidth]{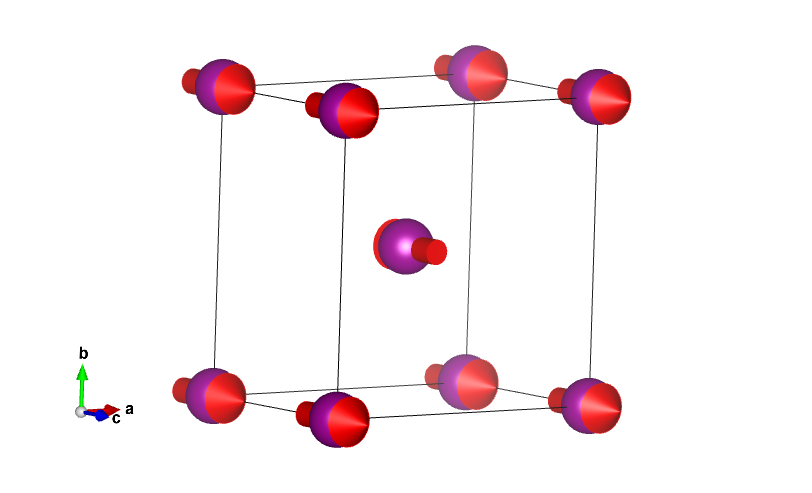}
	\caption{\label{fig_MnF}Magnetic structure of MnF$_2$ showing only the magnetic (Mn) atoms.}
\end{figure}

The magnetic phase of MnF$_2$ reported in \citet{Yamani2010} was assigned to the magnetic space group $P4_2^{\prime}/mnm^{\prime}$ (N. 136.499) in MAGNDATA \cite{Gallego2016}. The spin alignment presents a collinear structure with nontrivial SpPG $^{\overline{1}}4/\,^{1}m\,^{\overline{1}}m\,^{1}m$, and spin orientation direction along the (001) direction (see Fig. \ref{fig_MnF}) \cite{Chen2024}.
We choose the usual tetragonal crystallographic basis as the lattice and spin basis $({\bf a},{\bf b},{\bf c})=({\bf a}_s,{\bf b}_s,{\bf c}_s)$. Under this setting, the generators of the nontrivial SpPG can be written as:
\begin{equation}\begin{array}{ll}
	-y,x,z,-u,-v,-w &\{-1||4_{001}\}\\
	x,y,-z,u,v,w    &\{1||m_{001}\}\\
	-x,y,z,-u,-v,-w &\{-1||m_{100}\}.
\end{array}\end{equation}
The right-hand column displays the symmetry operations in a generalized Seitz notation. The spin orientation is introduced by selecting "collinear” along the (001) direction. For this spin orientation, the corresponding MPG is $4^{\prime}/mmm^{\prime}$. In this case, an alternative way to provide the input data using Option A is through the mcif file of the material since the SpPG is minimal. That file can be downloaded from MAGNDATA.

We now present some typical examples of tensor forms under the spin and magnetic symmetries. In all cases the tensors are expressed in an orthonormal basis with axes parallel to $({\bf a},{\bf b},{\bf c})$. For example, considering a [V$^2$]M tensor, which may represent the symmetric spin contribution $R_{ijk}^{(s)}$ to the Hall effect resistivity, we obtain the following symmetry-adapted form
\begin{equation}\label{eq:resist}\left(\begin{array}{ccc}
	0&0&0\\
	0&0&0\\
	0&0&0\\
	c_{41}&0&0\\
	0&c_{41}&0\\
	0&0&c_{63}
\end{array}\right)\end{equation}
for the MPG. The Hall effect resistivity connects the electric field ${\bf E}$ with the current density ${\bf J}$ and the magnetic field ${\bf H}$ according to $E_i=R_{ijk} J_j H_k$. $R_{ijk}^{(s)}$ is the symmetric part of $R_{ijk}$, $R_{ijk}^{(s)}=\frac{1}{2}\left(R_{ijk} +R_{jik}\right)$,  and accounts for the linear magnetoresistance (see \citet{Grimmer2017}). In Eq. (\ref{eq:resist}) we have used the Voigt notation contracting the first two indices into a single index which ranges from 1 to 6. In contrast, only $c_{63}$ remains nonzero under the SpPG symmetry, which means that $c_{41}$ must be a small effect arising from SOC. Similarly, the antisymmetric part of the spin contribution to the Hall effect, $R_{ijk}^{(a)}=\frac{1}{2}\left(R_{ijk}-R_{jik}\right)$, which is of type $a$\{V$^2$\}M, is entirely suppressed by the spin group symmetry.  However, under the MPG, two independent coefficients are allowed in the tensor, which is of the form
\begin{equation}\left(\begin{array}{ccc}
		0&0&0\\
		0&0&c_{123}\\
		0&c_{132}&0\\
		0&0&-c_{123}\\
		0&0&0\\
		-c_{132}&0&0\\
		0&-c_{132}&0\\
		c_{132}&0&0\\
		0&0&0
	\end{array}\right).\end{equation}
Therefore, in this case the entire property is expected to be a small relativistic effect.

Another example of a slightly different nature is the analysis of the spin splitting in energy bands. Among the predefined tensor properties available in STENSOR are the $(n+1)$-rank Cartesian tensors $T^{(n)}$, which describe the nonrelativistic spin splitting of electronic bands and spin textures. According to \citet{Radaelli2024}, the direction of the spin electronic polarization and the magnitude of the nonrelativistic spin splitting for a given band are given by the vector
\begin{equation}
	B_i^{eff} =\sum_{n=0}^{\infty}T_{i,\alpha,\beta,\ldots}^{(n)} k_{\alpha} k_{\beta}\ldots
\end{equation}
where $k_\alpha$  are the components of the wave vector, and the tensor coefficients depend on both the modulus of the wave vector and the band index. The Jahn symbols associated with $T^{(n)}$ are M[V$^n$] for even $n$ and $a$M[V$^n$] for odd $n$. For MnF$_2$, it is straightforward to verify that the lowest-order nonzero tensor is $T^{(2)}$, which contains a single surviving coefficient, $T_{321}^{(2)}=T_{312}^{(2)}$. This result directly implies that the material is a $d$-wave altermagnet \cite{Smejkal2022a,Smejkal2022b}, exhibiting non-relativistic spin splitting proportional to $k_x k_y$  to lowest order. Furthermore, since 
\begin{equation}
	B_1^{eff} =B_2^{eff}=0,\,B_3^{eff}\neq0 
\end{equation}
the electronic spin polarization is uniform to the lowest order across the entire Brillouin zone and is aligned with the  $z(c)$ direction. \subsection{Mn$_3$Sn (entry 0.199 in MAGNDATA)}
\label{sec:MnSn}
\begin{figure}[ht]
	\centering
	\includegraphics[width=0.8\textwidth]{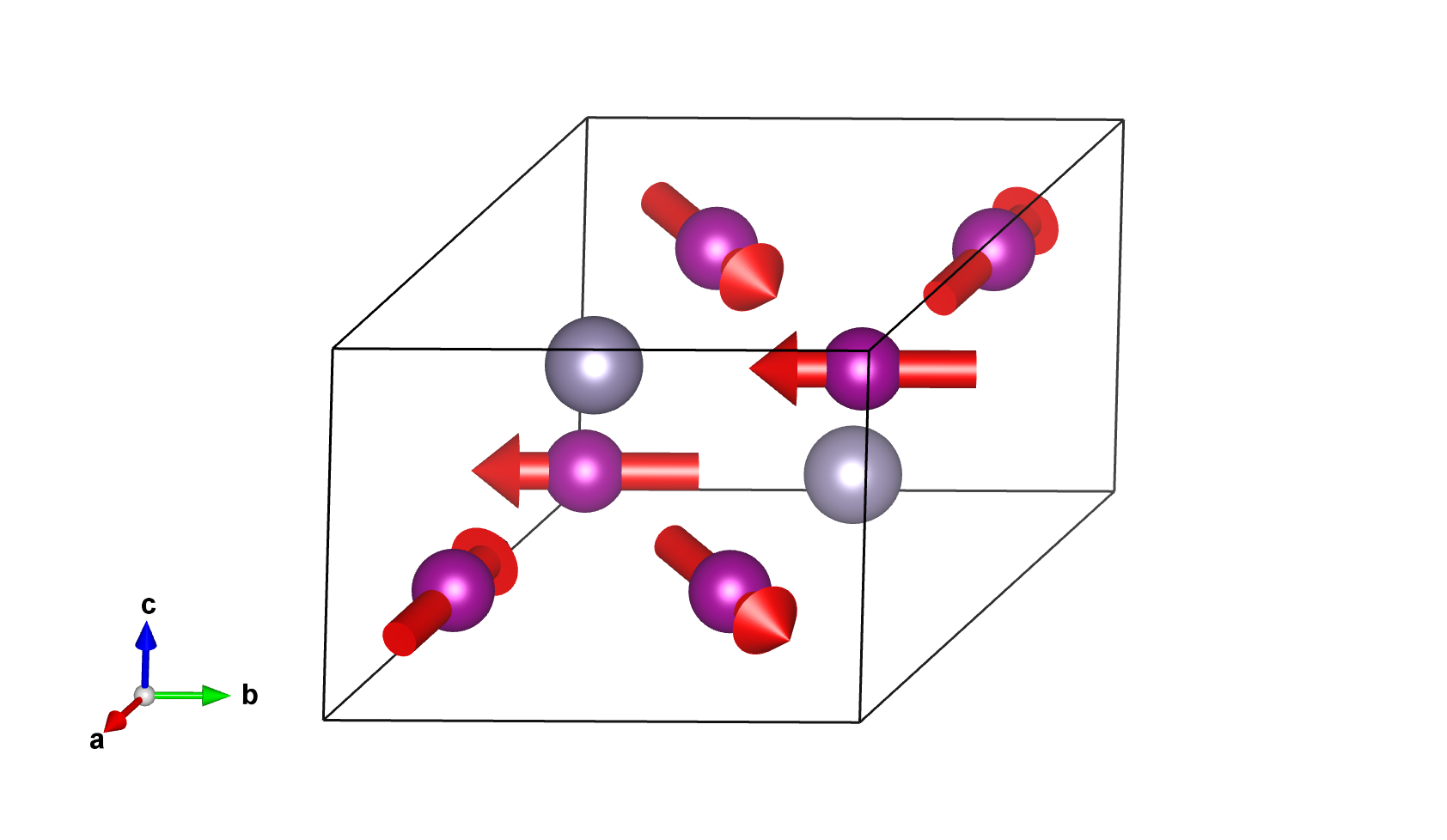}
	\caption{\label{fig_MnSn}Magnetic structure of Mn$_3$Sn showing the spins of the Mn atoms (entry 0.199 in MAGNDATA). Gray spheres represent the nonmagnetic Sn atoms. In the $({\bf b},-2{\bf a}-{\bf b},{\bf c})$ setting, the MPG is $mm^{\prime}m^{\prime}$.}
\end{figure}

One of the proposed structures for the magnetic phase of Mn$_3$Sn has magnetic group $Cmc^{\prime}m^{\prime}$ (N. 63.463) \cite{brown1990} and, according to \citet{Chen2024} the SpPG is given by
$^{3_{001}}6_{001}/\,^{1}m_{001}\,^{2_{110}}m_{100}\,^{2_{010}}m_{210}\,^{m_{001}}1$,  where the subindices refer to the hexagonal basis $({\bf a},{\bf b},{\bf c})$ (see Fig. \ref{fig_MnSn}).
Initially, we use the same reference frame both for space and spin operations.  The generators of the nontrivial SpPG can then be written as
\begin{equation}\begin{array}{ll}
       x-y,x,z,-v,u-v,w & \{3_{001}||6_{001}\}\\
		   x,y,-z,u,v,w & \{1||m_{001}\}\\
		-x+y,y,z,v,u,-w & \{2_{110}||m_{100}\}\\
   -x,-x+y,z,-u,-u+v,-w & \{2_{010}||m_{210}\}.
\end{array}\end{equation}
The type of spin alignment is in this case introduced in the program as “coplanar”, with the (001) direction perpendicular to the spin plane. Note that the “spin basis” option has not been used so far, similar to the previous example, since $({\bf a}_s,{\bf b}_s,{\bf c}_s)=({\bf a},{\bf b},{\bf c})$.

We can now examine the shapes of different simple representative tensors. In all cases the tensors are expressed in an orthonormal basis with axes parallel to $({\bf b},-2{\bf a}-{\bf b},{\bf c})$. The MPG is identified as the $mm^{\prime}m^{\prime}$ group in this same setting. 

For example, a M-type tensor like the spin magnetization is null according to the SpPG, but of the form $(c_1,0,0)$ for the MPG symmetry, indicating that the material can show weak ferromagnetism of SOC origin along the first axis. Similarly, [V$^2$] or [M$^2$] tensors — which correspond to electric and magnetic susceptibilities (orbital contribution), or spin magnetic susceptibility — are diagonal uniaxial, $c_{11}=c_{22}\neq c_{33}$, under the SpPG but diagonal biaxial under the MPG, $c_{11}\neq c_{22}\neq c_{33}$.

Finally, the tensor describing the anomalous Hall effect (AHE, type $a$\{V$^2$\}) is null for the SpPG, but results to be
\begin{equation}\left(\begin{array}{ccc}
		0&0&0\\
		0&0&c_{23}\\
		0&-c_{23}&0
\end{array}\right)\end{equation}
under the MPG, showing that a possible AHE in the material must be a SOC effect.

More complicated tensors can also be analyzed easily. For example, the spin contribution to the piezomagnetic tensor is of type M[V$^2$]. Under the MPG that tensor is
\begin{equation}\label{eq:piezo}\left(\begin{array}{cccccc}
		c_{11}&c_{12}&c_{13}&0&0&0\\
		0&0&0&0&0&c_{26}\\
		0&0&0&0&c_{35}&0
\end{array}\right),\end{equation}
where we have used the contraction of the last two indices, while the SpPG symmetry gives
\begin{equation}\label{eq:piezospin}\left(\begin{array}{cccccc}
		c_{11}&-c_{11}&0&0&0&0\\
		0&0&0&0&0&c_{11}\\
		0&0&0&0&0&0
\end{array}\right).\end{equation}
Since the $T^{(2)}$ tensor for spin textures shares the same Jahn symbol as the spin contribution to the piezomagnetic tensor, and $T^{(0)}=T^{(1)}=0$ for Mn$_3$Sn, Eq. (\ref{eq:piezospin}) implies that the lowest-order nonrelativistic spin splitting is described by the vector field:
\begin{equation}
B_1^{eff} =c_{11}\left(k_x^2-k_y^2\right),B_2^{eff} =2c_{11} k_x k_y,\,B_3^{eff} =0.
\end{equation}
These relations indicate that this coplanar material also exhibits $d$-wave magnetism, with the electronic spin polarization confined to the atomic spin plane.

The orbital contribution to the piezomagnetic tensor has the same form (\ref{eq:piezo}) under the MPG but is null under the SpPG.

\begin{figure}[ht]
	\centering
	\includegraphics[width=0.4\textwidth]{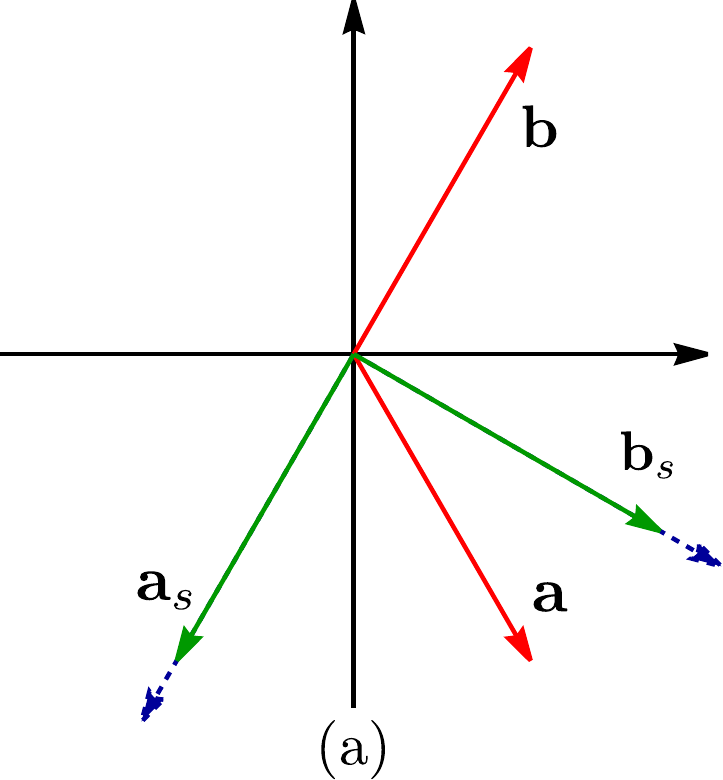}\hspace{2cm}\includegraphics[width=0.4\textwidth]{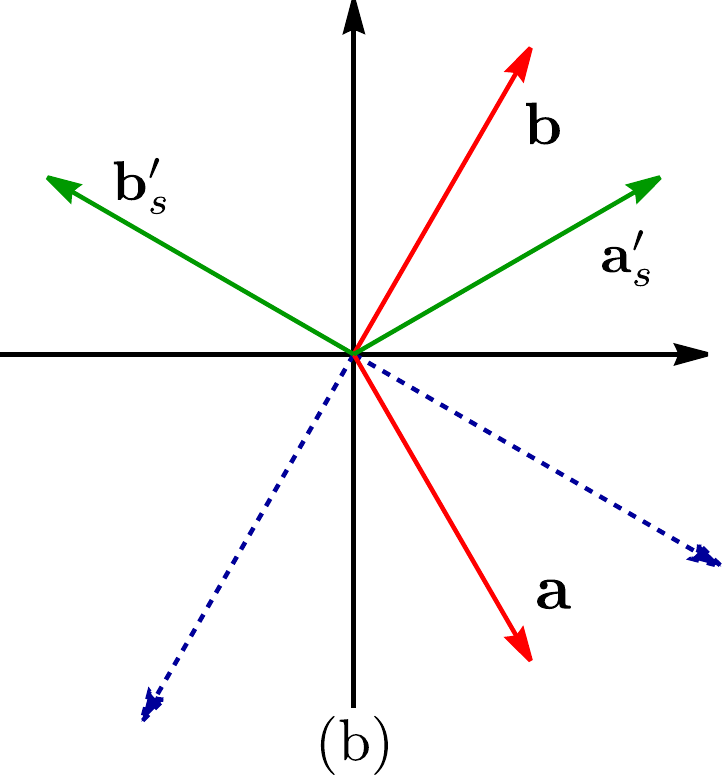}
	\caption{\label{fig_axes}Relationship between the hexagonal unit vectors ${\bf a}$ and ${\bf b}$ (red) of Mn$_3$Sn and the spin basis vectors (green). In (a) a Cartesian basis, with vectors ${\bf a}_s$ and ${\bf b}_s$, is used. In (b) an oblique spin basis, with vectors ${\bf a}_s^{\prime}$ and ${\bf b}_s^{\prime}$, is employed to describe the Mn$_3$Sn structure with the spins rotated by 90$^{\circ}$. In both figures the blue dotted lines represent the first and second axes of the reference frame used to express the tensors.}
\end{figure}

We now illustrate the use of the “spin basis” option by analyzing the same material but adopting this time a different reference frame for the spin space. For example, if we take an orthonormal spin basis given by (see Fig. \ref{fig_axes}a)
\begin{equation}\label{eq:sqrt1}\begin{array}{rcl}
	 {\bf a}_s&=&-{\bf b}\\
	 {\bf b}_s&=&\frac{2}{\sqrt{3}}{\bf a}+\frac{1}{\sqrt{3}}{\bf b}\\
	 {\bf c}_s&=&{\bf c},
\end{array}\end{equation}
the generators can be expressed as
\begin{equation}\label{eq:sqrt2}\begin{array}{ll}
		x-y,x,z,-\frac{1}{2}u-\frac{\sqrt{3}}{2}v,\frac{\sqrt{3}}{2}u-\frac{1}{2}v,w & \{3_z||6_{001}\}\\
		x,y,-z,u,v,w & \{1||m_{001}\}\\
		-x,-x+y,z,u,-v,-w & \{2_x||m_{210}\}.
\end{array}\end{equation}
where the subindices $x$ and $z$ give the orientation of the 2- and 3-fold axes with reference to the spin basis. Evidently, the same results are obtained in the end for the symmetry-adapted tensors. In the program, the required format for the square roots is sqrt(.), both for the basis relations [Eqs. (\ref{eq:sqrt1})] and operations [Eqs. (\ref{eq:sqrt2})].

\begin{figure}[ht]
	\centering
	\includegraphics[width=0.8\textwidth]{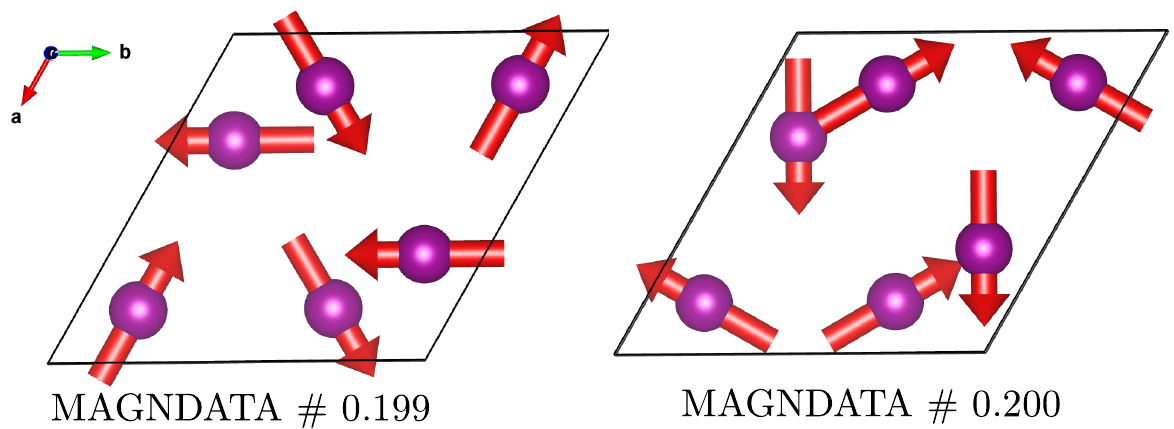}
	\caption{\label{fig_MnSn-0-200}Two inequivalent structures of Mn$_3$Sn showing only the spins of the magnetic atoms. The second structure is obtained from the first one by rotating the spins 90$^{\circ}$ counterclockwise about the ${\bf c}$ axis. In the $({\bf b},-2{\bf a}-{\bf b},{\bf c})$ setting, the MPGs are $mm^{\prime}m^{\prime}$ and $m^{\prime}mm^{\prime}$, respectively.}
\end{figure}

The “spin basis” option goes beyond the mere possibility of creating an orthogonal reference frame for the spin coordinates. In addition to the structure shown in Fig. \ref{fig_MnSn}, another model has also been reported for Mn$_3$Sn \cite{brown1990}, which experimentally could not be distinguished from the former. In this second model the spins are rotated by 90$^{\circ}$ counterclockwise around the $c$ axis with respect to the spins in the first model (see Fig. \ref{fig_MnSn-0-200}). However, both structures have different MSGs and are not physically equivalent. Thus, there are two different entries (0.199 and 0.200) in MAGNDATA. From the viewpoint of spin group symmetry, the two structures have however the same SpSG, as the only difference is a rotation of the spin arrangement. The spin operations $U$ have however a different orientation with respect to the lattice, and this implies in general different tensor properties. Hence, the oriented SpPG of this second structure is $^{3_{001}}6_{001}/\,^{1}m_{001}\,^{2_{1\overline{1}0}}m_{100}\,^{2_{210}}m_{210}\,^{m_{001}}1$ and, in this basis, the generators to be introduced for the nontrivial SpPG would be different. However, by using the spin basis option, we can work with the same input operations as in the previous case (0.199), and instead, simply refer the spin operations of the input generators to a spin basis rotated by 90$^{\circ}$ around the $c$ axis. The new spin basis is then (see Fig. \ref{fig_axes}b)
\begin{equation}\begin{array}{rcl}
		{\bf a}_s^{\prime}&=&\frac{1}{\sqrt{3}}{\bf a}+\frac{2}{\sqrt{3}}{\bf b}\\
		{\bf b}_s^{\prime}&=&-\frac{2}{\sqrt{3}}{\bf a}-\frac{1}{\sqrt{3}}{\bf b}\\
		{\bf c}_s^{\prime}&=&{\bf c},
\end{array}\end{equation}

In this way, the former directions $(110)$ and $(010)$ in the spin space are now $(1\overline{1}0)$ and $(210)$, respectively, while the $(001)$ direction remains invariant. Thus, the SpPG is correctly oriented for the structure of entry 0.200, with the corresponding MPG being now $m^{\prime}mm^{\prime}$. There are several instances where it can be seen that the rotated structure is not equivalent to the former. For example, it can be checked that the new symmetry allows weak ferromagnetism but with the magnetization along the second axis $(-2{\bf a}-{\bf b})$ instead of the first one (${\bf b}$), since an M-type tensor vanishes under the SpPG, but takes the form $(0,c_2,0)$ under the MPG. Similarly, Eqs. (\ref{eq:piezo}) and (\ref{eq:piezospin}) become
\begin{equation}\left(\begin{array}{cccccc}
		0&0&0&0&0&c_{16}\\
		c_{21}&c_{22}&c_{23}&0&0&0\\
		0&0&0&c_{34}&0&0
	\end{array}\right)\end{equation}
and
\begin{equation}\left(\begin{array}{cccccc}
		0&0&0&0&0&c_{16}\\
		-c_{16}&c_{16}&0&0&0&0\\
		0&0&0&0&0&0
	\end{array}\right),\end{equation}
respectively for the structure of entry 0.200. The change of the tensors can be rationalized as a mere 90$^{\circ}$ rotation of the spin degrees of freedom of the tensors with respect to the lattice. But, as directions differing by 90$^{\circ}$ in an hexagonal system are not equivalent, this difference is physically relevant. \section{Methods and technical details}
\label{sec:methods-main}
STENSOR is a program written in the \emph{Mathematica} language (Wolfram). Below, we provide a brief description of the mathematical algorithms used in the tensor reduction process, the general functioning of the program, and several relevant technical details. Additional information can be found in the Supporting Information.

The program checks first that the input data introduced have a correct format. If Option A is used, it is assumed that the scif or mcif files are properly constructed. The program then simply verifies that the Jahn symbol is correct. In Option B, in addition to this check, the program confirms that the matrix relating the spin and space bases is non-singular, and that at least one of the three components ${\bf n}=(n_1,n_2,n_3)$ of the direction parallel (perpendicular) to the distributions of spins in collinear (coplanar) structures is non-zero. The program accepts float numbers as components of ${\bf n}$ but they are immediately transformed as a set of integer numbers that represent the same direction. Every row in the box of generators must correspond to a pair of matrices $\{U||R\}$ that are non-singular. Once the matrices have been constructed, it is checked that successive multiplications of the generators end in a finite number of symmetry operations so that the generated elements form a finite point group.

The calculation is divided in several steps, described in the following subsections.
 \subsection{Determination of the magnetic point group as a subgroup of the spin point group}
\label{sec:methods-main-MPG}
Once all the operations $\{U||R\}$ of the SpPG associated with a non-trivial SpSG have been determined—based on the scif or mcif files (Option A) or the introduced set of generators (option B)—those elements within the SpPG that also belong to the MPG are identified. Provided that the spin and lattice bases coincide, these elements are of the form $\{\pm R||R\}$. If the two bases differ, the identification is carried out by first performing a change of basis in the spin space.

As demonstrated in Section \ref{sec:determination-MPG} of the Supporting Information, in the collinear and coplanar cases, an efficient algorithm to identify such operations is through the condition $U^{-1}R{\bf n}=\pm{\bf n}$. For noncoplanar groups the identification is straightforward. If a mcif file is used in Option A, a minimal SpPG is assumed and the MPG operations are explicit.

Once such operations are listed, the program determines the corresponding MPG. Since the operations are generally not expressed in the standard setting of the identified MPG \cite{Litvin2013}, STENSOR provides a transformation matrix $P$ that changes the original setting to the standard one. This transformation satisfies the relation $R_s=P^{-1}RP$, where $R_s$ is the matrix representation of the symmetry operation in the standard setting. \subsection{Tensor reduction under the magnetic point group}
\label{sec:main-mpgreduction}
The tensor reduction under the MPG is performed using projectors \cite{bradley1972,dresselhaus2008}. The procedure is carried out in two steps. In the first step, possible symmetry or antisymmetry between pairs of tensor indices is not considered, and only the point group operations are applied. In the second step, the additional constraints imposed by the (anti)symmetry of the tensor indices are incorporated. The procedure involves working with matrices $3^r\times3^r$, where $r$ is the rank of the tensor to be reduced. To simplify the handling of these matrices STENSOR performs a row reduction (or Gauss decomposition) to convert them into the so-called reduced row echelon forms. Additionally, the {\bf SparseArray} functionalities implemented in \emph{Mathematica} are used to efficiently store and operate on large matrices with a small fraction of nonzero elements. These options greatly improve the memory and cpu-time requirements in the calculations, and are especially useful for finding the symmetry adapted forms of tensors with large ranks. A detailed explanation of the method is provided in the Supplementary Information (Sections \ref{sec:reductionMPG}, \ref{sec:intrinsic} and \ref{sec_extra-considerations}). \subsection{Tensor reduction under the spin point group}
\label{sec:main-spgreduction}
The calculation of the tensor reduction under the full spin group can be divided into three steps, with the first two being almost identical to those used for the reduction under the MPG. First, the constraints imposed by all operations of the SpPG, except those belonging to the intrinsic spin-only group, are determined. Next, the symmetry constraints arising from the symmetry or antisymmetry of the tensor indices are incorporated. Finally, as a third and final step, the additional reduction imposed by the intrinsic spin-only point group in collinear and coplanar spin arrangements is applied. Here as well, the entire procedure employs the projector technique (except for the infinite-fold axis in the collinear spin-only group), Gaussian decomposition, and the {\bf SparseArray} method.

In the case of the collinear spin-only group, the order of the group $\infty_{{\bf n}}$ is infinite, and it is not possible to define a projector; therefore, a different procedure must be followed. This infinite group can be regarded as generated by an infinitesimal rotation about ${\bf n}$ by an angle $\varphi\ll1$. Expanding the matrix representing such a rotation as a series and retaining only the linear term, one can show that the invariance condition under $\infty_{{\bf n}}$ leads to a set of homogeneous linear equations. These equations impose linear relations among the independent tensor coefficients, thereby completing the tensor reduction. 
The entire procedure is thoroughly explained in Sections \ref{sec:reduction-spg} and \ref{sec:reduction-trivial-group} of the Supporting Information. In addition, an explicit example illustrating all steps of the method in detail is provided in Section \ref{sec:example}. \section{Conclusions}
\label{sec:conclusions}
This study has presented STENSOR, a new addition to the Bilbao Crystallographic Server, which automates the derivation of symmetry-adapted tensor forms under SpPG symmetry. The tool can operate from either structural files (scif or mcif) or from manually provided generators of the oriented SpPG in combination with a generalized Jahn symbol, producing as output the tensor forms allowed by both the SpPG and MPG. This double output makes it straightforward to identify which tensor components have nonrelativistic origin and which ones typically arise from SOC. From a computational point of view, STENSOR is implemented in Mathematica, and relies on algorithms such as the projector method, Gaussian reduction, and SparseArray-handling to reduce memory and cpu requirements. Tensors of ranks as high as 6 or 7 can be processed efficiently, with computation times remaining reasonable.

The case of two real materials, collinear MnF$_2$ and coplanar Mn$_3$Sn, has been analyzed to illustrate some of the STENSOR capabilities in determining the forms of representative tensors. In MnF$_2$ STENSOR finds a $d$-wave altermagnetic spin splitting ($k_xk_y$ dependence) with uniform electronic spin polarization along the $z$ axis. For Mn$_3$Sn, $d$-wave magnetism is also identified, but here the spin polarization is confined on the $x-y$ plane. In this compound, it is also shown how the spin-basis option can be used to describe structures which differ in a rigid spin rotation. 
STENSOR complements existing tools such as MTENSOR and databases like MAGNDATA within the Bilbao Crystallographic Server. The program is expected to be useful to distinguish SOC-driven versus nonrelativistic effects, and may facilitate systematic studies in magnetic compounds.

We conclude with a note of caution: experimental magnetic structures may incorporate features that originate from SOC, thereby reducing the actual SpPG symmetry with respect to the ideal SOC-free case. In such situations, the comparison of tensor forms under MPG and SpPG do not separate properly relativistic from nonrelativistic contributions. In fact, tensor components allowed under the observed (reduced) SpPG may in practice require nonzero SOC. 
\acknowledgments
This work has been supported by the Government of the Basque Country (Project No. IT1458-22).

\onecolumngrid
\renewcommand{\thesection}{Appendix \arabic{section}}
\renewcommand{\thesubsection}{\arabic{section}.\arabic{subsection}}

\clearpage
\begin{center}
{\bf Supplementary Material of "Automatic calculation of symmetry-adapted tensors under spin-group symmetry. STENSOR, a new tool of the Bilbao Crystallographic Server."}
\end{center}

\tableofcontents

\clearpage

\addtocontents{toc}{\protect\setcounter{tocdepth}{3}}
\addtocontents{lot}{\protect\setcounter{lotdepth}{3}}
\renewcommand{\thetable}{S\arabic{table}}
\renewcommand{\thefigure}{S\arabic{figure}}
\renewcommand{\thesection}{S\arabic{section}}
\renewcommand{\thesubsection}{\thesection.\arabic{subsection}}
\renewcommand{\thesubsubsection}{\thesubsection.\arabic{subsubsection}}

\makeatletter
\def\p@subsection{}
\def\p@subsubsection{}
\makeatother

\setcounter{section}{0} 
\setcounter{table}{0} 
\setcounter{equation}{0} 

\section{Methods and technical details}
\label{sec:methods}
In Option A, all the required data except the Jahn symbol are obtained from the uploaded scif or mcif files. In a scif file the symmetry operations have the form $x,y,z,u,v,w$ or $x,y,z,+1,u,v,w$ where $x,y,z$ refer to the $R$ orbital or lattice part of the $\{U||R\}$ spin point group operation and $u,v,w$ refer to the spin part $U$. The integer $\pm1$ between both triplets in the second case is redundant and its value is $\det(U)$. The symmetry operations are converted into the matrix form $\{U||R\}$. In an mcif file the symmetry operations have the usual form of the symmetry operations of a magnetic point group, $x,y,z,\pm1$. As in the previous case, $x,y,z$ refer to the orbital part $R$ and the $U$ matrix can be obtained through the relation $U=\pm1\det(R)R$. The operations are converted to matrix form $\{U||R\}$. If the uploaded file is a scif file, in the next step the program reads the matrix that relates the bases in the orbital and spin spaces. If the uploaded file is an mcif file, by construction, both sets of matrices are expressed in the setting of the orbital space and the matrix is assumed to be the identity. Next the program checks whether the spin distribution is collinear, coplanar or non-coplanar, using the information in the file about the magnetic moments of the independent atoms in the asymmetric unit. First all the symmetry operations are applied to these independent moments to get the magnetic moments of all the atoms in the unit cell. If all spins are parallel or antiparallel the structure is collinear and the vector ${\bf n}$ is the direction parallel to the spins. If all the moments lie in the same plane the structure is coplanar and ${\bf n}$ is calculated through the cross product of the (non-parallel and not anti-parallel) moments of two atoms. Finally, the program checks that the Jahn symbol introduced by the user has the right format.
 
In Option B all the data are introduced manually by the user and the program checks first that these data have the right format: the Jahn symbol is correct, the $M$ matrix that relates the bases $({\bf a},{\bf b},{\bf c})$ and $({\bf a}_s,{\bf b}_s,{\bf c}_s)$ is non-singular and at least one of the three components ${\bf n}=(n_1,n_2,n_3)$ of the direction parallel (perpendicular) to the distributions of spins in collinear (coplanar) structures is non-zero. The program accepts float numbers as components of ${\bf n}$ but they are immediately transformed as a set of integer numbers that represent the same direction.

 Every row in the box of generators must correspond to a pair of matrices $\{U||R\}$ that are non singular. Once the matrices have been constructed, it is checked that successive multiplications of the generators end in a finite number of symmetry operations so that the generated elements form a finite point group. In the present form of the program, whereas the point group of the operations $R$ in the orbital space must be one of the 32  crystallographic point groups, the point group of the spin operations $U$ can be not only one of the 32 crystallographic point groups, but also a finite point group that contains a 8-fold or a 12-fold proper or improper rotation (see Table \ref{tab:noncryst} for a complete list of allowed non crystallographic point groups). Moreover, the program checks whether all the $U$ operations keep invariant (or reverse) ${\bf n}$ in collinear or coplanar cases.

\begin{table}[h]
	\caption{\label{tab:noncryst} List non-crystallographic point groups of the $U$ operations allowed by the input of STENSOR.}
	\begin{center}
		\begin{tabular}{lllllll}
			8&$\overline{8}$&$8/m$&822&$8mm$&$\overline{8}m2$&$8/mmm$\\
			12&$\overline{12}$&$12/m$&12\,22&$12mm$&$\overline{12}m2$&$12/mmm$
		\end{tabular}
	\end{center}
\end{table}

Once that the parameters of the input have been accepted, and the whole set of symmetry operations of the non trivial group have been determined, the $U$ matrices, expressed initially in the $({\bf a}_s,{\bf b}_s,{\bf c}_s)$ basis, are transformed into the setting of the orbital space through the given $M$ matrix, $U'=MUM^{-1}$. If the user has not made use of this option, it is understood by the program that both $R$ and $U$ are expressed in the same basis and, then, the $M$ matrix is the identity. In the collinear and coplanar cases the components of ${\bf n}$ are also transformed as ${\bf n}'=M{\bf n}$. Therefore, in the calculation of the tensor reduction, $U'$ (from now on $U$) and $R$ are expressed in the same reference system.

The calculation is divided in several steps, described in the following sections. \subsection{Determination of the magnetic point group as a subgroup of the spin point group}
\label{sec:determination-MPG}
As has been stressed in the main text, it is interesting to compare the form of a tensor under the given SpPG and under its subgroup whose operations form a maximal MPG. We have thus to identify those operations $\{U||R\}$ that satisfy $U=\pm R$ and that compose the MPG \cite{etxebarria2025}.

The operations of the point group can be expressed, in general, as the direct product of the nontrivial SpPG, \Pnt\, and the spin-only group \Pso\,,
	\begin{equation}
	\label{e-Pss-descom}
	\textrm{P}_{\textrm{\footnotesize{S}}} = \textrm{P}_{\textrm{\footnotesize{NT}}} \times \textrm{P}_{\textrm{\footnotesize{SO}}},
\end{equation}
where \Pso\, contains all operations of the form $\{U||1\}$ and can, in turn, be expressed as the direct product of the \emph{intrinsic} or trivial \Psoin\, point group and \Psog\,
	\begin{equation}
	\label{e-Pso-descom}
	\textrm{P}_{\textrm{\footnotesize{SO}}} = \textrm{P}_{\textrm{\footnotesize{SOG}}} \times \textrm{P}_{\textrm{\footnotesize{SOintr}}}.
\end{equation}
The intrinsic group \Psoin\, is the identity in non-coplanar distributions of spins, $^{m_{\bf n}}1$ in coplanar distributions, where $m_{\bf n}$ represents a mirror plane whose normal is parallel to ${\bf n}$ and $^{\infty_{\bf n}m}1$ in collinear distributions, where $\infty_{\bf n}$ is the rotation axis of order infinite parallel to ${\bf n}$. \Psog\, contains, together with the identity, all the spin-only operations that do not belong to the trivial group.
STENSOR splits the point group into the intrinsic part \Psoin\, on one hand, and the rest of terms in the decomposition of \Ps. This group,\Pnte, is the SpPG associated to the nontrivial SpSG, i.e., \Pnte=\Pnt$\times$\Psog. Using this split of the point group the determination of the MPG is straightforward in the three cases of spin arrangements.

In collinear groups we take every operation of $\{U||R\}$ in \Pnte\, and check whether there exits some operation $\{U'||1\}$ in \Psoin:$^{\infty_{\bf n}m}1$ such that $UU'=\pm R$. However, by construction, $^{\infty_{\bf n}m}1$ is the set of symmetry operations that keep ${\bf n}$ invariant, it is sufficient just to check whether the product $\pm U^{-1}R$ belongs to the intrinsic group, i.e., whether one of the following two conditions is fulfilled,
\begin{equation}
	\label{eq:condcollinear}
	U^{-1}R{\bf n}=\theta{\bf n},
\end{equation}
with $\theta=\pm1$. If the condition  is fulfilled for $\theta$, the symmetry operation ${\{R,\theta\det(R)\}}$ belongs to the MPG, being a unitary operation if $\theta\det(R)=1$ and anti-unitary if $\theta\det(R)=-1$. 

In coplanar systems the procedure is very similar. The operation $\{U||R\}$ belongs to the MSG if there exits some operation $U'$ in the intrinsic point group (that contains only the identity and a mirror plane) such that $\theta U^{-1}R=U'$. This condition can be divided into two:
\begin{itemize}
	\item $\theta U^{-1}R=1$ with $\theta=\pm1$. 
	\item $\theta U^{-1}R=m_{\bf n}$ which is fulfilled if $(U^{-1}R)^2=1$, $\theta U^{-1}R\neq-1$, $\det(\theta U^{-1}R)=-1$ and  \mbox{$\theta U^{-1}R{\bf n}=-{\bf n}$}.
\end{itemize}
In both cases the operation $\{R,\theta\det(R)\}$ belongs to the MPG.

In non-coplanar cases, as the intrinsic group is the identity, we have to select those operations $\{U||R\}$ of \Pnte\, that fulfill the relation $U=\theta R$. The operation $\{R,\theta\det(R)\}$ belongs to the MPG.

Once all the operations of the MPG have been listed, the program identifies the magnetic point group. As  the operations will not be expressed in general in the standard setting of the identified MPG, together with the symbol and number of the MPG, STENSOR provides a transformation matrix $P$ from the setting where the original $R$ operations are expressed to the standard setting of the MPG \cite{Litvin2013} such that $R_s=P^{-1}RP$, being $R_s$ the matrix of the symmetry operation in the standard setting.
 
 \subsection{Tensor reduction under the magnetic point group}
\label{sec:mpgreduction}
In this section we describe the algorithm used by STENSOR to calculate the constrains on a given tensor imposed by the magnetic point group. 
We can distinguish two steps in the tensor reduction. In the first step the (anti)symmetrization of the tensor is not considered, so only the point group operations are used and, in the second step, we add the conditions imposed by the (anti)symmetric subsets of components enclosed into $[]$ and $\{\}$ symbols. 
\subsubsection{Tensor reduction by the operations of the MPG}\label{sec:reductionMPG}
If no (anti)symmetrization of indices is considered, one takes thus a tensor of rank $r$ whose Jahn symbol (once the symmetrization and/or anti-symmetrization of some of its components have been removed) is V$r$, $a$V$r$, $e$V$r$ or $ae$V$r$. 

Let $\{R^i,\theta^i\}$ be an element of the MPG with $\theta^i=1,-1$ for unitary and anti-unitary operations, respectively, and let $T_{i_1,i_2,\ldots,i_r}$ be the components of a tensor of rank $r$, with $i_j=1,2,3$ for all $j=1,\ldots,r$. Any element of the MPG introduces a set of restrictions on these components given by
	\begin{equation}
		\label{eq:3dimreduction}
		T_{i_1,i_2,\ldots,i_r}=T_{i_1',i_2',\ldots,i_r'}=f^iR_{i_1'i_1}^iR_{i_2'i_2}^i\ldots R_{i_r'i_r}^sT_{i_1,i_2,\ldots,i_r},
	\end{equation}
with $f^i=1$, $\theta^i$, $\det(R^i)$ or $\theta^i\det(R^i)$ for tensors of type V$r$, $a$V$r$, $e$V$r$ or $ae$V$r$, respectively.  We now define a $3^r$-dimensional vector ${\bf T}$ whose $u$ component is $T_u=T_{i_1,i_2,\ldots,i_r}$ such that
\begin{equation}
	\label{eq:ordercomponents}
	u=3^{r-1}(i_1-1)+3^{r-2}(i_2-1)+\ldots+3(i_{r-1}-1)+(i_r-1)+1.
\end{equation}
The vector ${\bf T}$ contains thus all the $3^r$ components of the tensor ordered in a specific way.
It is straightforward to check that the relation (\ref{eq:3dimreduction}) transforms into
\begin{equation}
	\label{eq:linearequations}
	T_u=T_{u'}=\mathcal{R}_{u'u}^iT_u,
\end{equation}
being $\mathcal{R}^i$ the Kronecker (or direct) product of $r$ matrices $R^i$ times $f^i$,
\begin{equation}
	\label{eq:Kronecker}
	\mathcal{R}^i=f^i\overbrace{R^i\otimes R^i\otimes\ldots\otimes R^i}^r.
\end{equation}

It is possible to stablish group isomorphisms between the point group operations $R^i$, the MPG operations $\{R^i|\theta^i\}$ and the $\mathcal{R}^i$ matrices such that,
\begin{equation}\begin{array}{rcrcl}
	R^1&\rightarrow&\{R^1|\theta^1\}&\rightarrow&\mathcal{R}^1=f^1\overbrace{R^1\otimes R^1\otimes\ldots\otimes R^1}^r,\\
	R^2&\rightarrow&\{R^2|\theta^2\}&\rightarrow&\mathcal{R}^2=f^2\overbrace{R^2\otimes R^2\otimes\ldots\otimes R^2}^r.
\end{array}\end{equation}
If
\begin{equation}
	\label{eq:product}
		\{R^3|\theta^3\}=\{R^1|\theta^1\}\{R^2|\theta^2\}=\{R^3=R^1R^2|\theta^3=\theta^1\theta^2\},
\end{equation}
making use of the properties of the Kronecker product, it is immediate to check that,
\begin{equation}
	\mathcal{R}^3=\mathcal{R}^1\mathcal{R}^2.
\end{equation}
It is important to note that equation (\ref{eq:product}) represents the product of symmetry operations without conjugation when $\theta^i=-1$. The consequences on the tensor reduction of the anti-unitary operations is carried out by the multiplication by $\theta^i$ in equation (\ref{eq:3dimreduction}), when the Jahn symbol contains the $a$ factor \cite{Grimmer1993,Grimmer2017,Gallego2019}.
Therefore, the $\mathcal{R}^i$ matrices of dimension $3^r\times3^r$ form a $3^r$ dimensional representation $\rho$ of the point group $\mathcal{P}$ formed by the $R^i$ operations. The determination of the number of independent components of the vector ${\bf T}=(T_1,\ldots,T_{3^r})^T$ reduces thus to calculate the multiplicity of the identity representation in $\rho$, and the calculation of a set of basis vectors that transform under the unitary irreducible representation can be easily performed making use of the projector operators \cite{bradley1972,dresselhaus2008}. If we consider the identity irreducible representation, the projector reduces to,
\begin{equation}
	\label{eq:projector}
	(\mathds{I}_{3^r})_{u'u}=\frac{1}{|\mathcal{P}|}\sum_{s=1}^{|\mathcal{P}|}\mathcal{R}^s_{u'u},
	\end{equation}
	where $\mathds{I}_{3^r}$ is the $3^r\times3^r$ identity matrix and $|\mathcal{P}|$ is the order of the point group $\mathcal{P}$. If one applies this projector into the vector that contains the tensor components the result is:
	\begin{equation}
		\label{eq:pointreduction}
		{\bf T}=\frac{1}{|\mathcal{P}|}\sum_{s=1}^{|\mathcal{P}|}\mathcal{R}^s{\bf T}\equiv \textrm{P}_{\mathcal{P}}{\bf T}.
	\end{equation}
	Thus, the number of independent components of the tensor is the rank $rank(\textrm{P}_{\mathcal{P}})$ of the matrix 
	\begin{equation}
		\label{eq:pointgroupreduction}
		\textrm{P}_{\mathcal{P}}=\frac{1}{|\mathcal{P}|}\sum_{s=1}^{|\mathcal{P}|}\mathcal{R}^s,
	\end{equation}
	i.e., the number of linearly independent rows of $\textrm{P}_{\mathcal{P}}$. Each row gives a relation between different components of the tensor although not all the rows are independent. If the tensor has no additional restrictions that come from the (anti)symmetry under the interchange of indices, this matrix has all the information about the reduced form of the tensor.
\subsubsection{Further tensor reduction under (anti)symmetrization of indices}
	\label{sec:intrinsic}
	Once the restrictions of the point group have been considered, in the second step the extra restrictions due to the (anti)symmetrization of indices are also taken into account.
	
	Let the tensor be symmetric (or antisymmetric) under the interchange of the $i_k$ and $i_{k+1}$ components, i.e. the Jahn symbol of the symmetric tensor is V$(k-1)$[V2]V$(r-k-1)$ (or V$(k-1)$\{V2\}V$(r-k-1)$ in the antisymmetric case) where the parentheses have been added for clarity. In the  $3^r$ dimensional space the $r$-rank tensor is described as ${\bf T}=(T_1,\ldots,T_{3^r})^T$ and the symmetric (antisymmetric) interchange of the components $i_k$ and $i_{k+1}$ is expressed through the following $\mathcal{S}$ ($\mathcal{A}$) matrices,
	\begin{equation}
		\label{eq:symantisym}
		\begin{array}{rcl}
		\mathcal{S}=\overbrace{\mathds{I}_3\otimes \mathds{I}_3\otimes\ldots\otimes \mathds{I}_3}^{k-1}\otimes S\otimes\overbrace{\mathds{I}_3\otimes \mathds{I}_3\otimes\ldots\otimes \mathds{I}_3}^{r-k-1},\\
		\mathcal{A}=\overbrace{\mathds{I}_3\otimes \mathds{I}_3\otimes\ldots\otimes \mathds{I}_3}^{k-1}\otimes A\otimes\overbrace{\mathds{I}_3\otimes \mathds{I}_3\otimes\ldots\otimes \mathds{I}_3}^{r-k-1},
	\end{array}\end{equation}
	with
	\begin{equation}
		\label{eq:SandAmatrices}
		S=\left(
		\begin{array}{ccccccccc}
			1 & 0 & 0 & 0 & 0 & 0 & 0 & 0 & 0 \\
			0 & 0 & 0 & 1 & 0 & 0 & 0 & 0 & 0 \\
			0 & 0 & 0 & 0 & 0 & 0 & 1 & 0 & 0 \\
			0 & 1 & 0 & 0 & 0 & 0 & 0 & 0 & 0 \\
			0 & 0 & 0 & 0 & 1 & 0 & 0 & 0 & 0 \\
			0 & 0 & 0 & 0 & 0 & 0 & 0 & 1 & 0 \\
			0 & 0 & 1 & 0 & 0 & 0 & 0 & 0 & 0 \\
			0 & 0 & 0 & 0 & 0 & 1 & 0 & 0 & 0 \\
			0 & 0 & 0 & 0 & 0 & 0 & 0 & 0 & 1 \\
		\end{array}
		\right)\hspace{2cm}A=\left(
		\begin{array}{ccccccccc}
			-1 & 0 & 0 & 0 & 0 & 0 & 0 & 0 & 0 \\
			0 & 0 & 0 & -1 & 0 & 0 & 0 & 0 & 0 \\
			0 & 0 & 0 & 0 & 0 & 0 & -1 & 0 & 0 \\
			0 & -1 & 0 & 0 & 0 & 0 & 0 & 0 & 0 \\
			0 & 0 & 0 & 0 & -1 & 0 & 0 & 0 & 0 \\
			0 & 0 & 0 & 0 & 0 & 0 & 0 & -1 & 0 \\
			0 & 0 & -1 & 0 & 0 & 0 & 0 & 0 & 0 \\
			0 & 0 & 0 & 0 & 0 & -1 & 0 & 0 & 0 \\
			0 & 0 & 0 & 0 & 0 & 0 & 0 & 0 & -1 \\
		\end{array}
		\right)
	\end{equation}
	and $\mathds{I}_3$ being the $3\times3$ identity matrix.
	
	The two matrices $\left(\mathds{I}_{3^r},\mathcal{S}\right)$ (or $\left(\mathds{I}_{3^r},\mathcal{A}\right)$) form a group of order 2 so that, on the reduced form of the tensor components $(T_1,T_2,\ldots,T_{3^r})^T$ calculated in the first step and developed in section \ref{sec:reductionMPG}, one can apply the same arguments. First, it is possible do define the projectors,
	\begin{equation}
		\mathds{I}_{3^r}=\frac{1}{2}\left(\mathds{I}_{3^r}+\mathcal{S}\right)\hspace{1cm}\textrm{and}\hspace{1cm}\mathds{I}_{3^r}=\frac{1}{2}\left(\mathds{I}_{3^r}+\mathcal{A}\right)
	\end{equation}
	in the symmetric and anti-symmetric case, respectively. The basis vectors can be calculated through the relations,
		\begin{equation}
			\label{eq:simanti}
		{\bf T}=\frac{1}{2}\left(\mathds{I}_{3^r}+\mathcal{S}\right){\bf T}\equiv \textrm{P}_{\mathcal{S}}{\bf T}\hspace{1cm}\textrm{and}\hspace{1cm}{\bf T}=\frac{1}{2}\left(\mathds{I}_{3^r}+\mathcal{A}\right){\bf T}\equiv\textrm{P}_{\mathcal{A}}{\bf T}.
	\end{equation}
	Using the relation (\ref{eq:pointreduction}) the symmetry reduction due to the MPG and the intrinsic symmetry of the tensor can be written as,
	\begin{equation}
		\label{eq:all}
		{\bf T}=\textrm{P}_{\mathcal{P}}\textrm{P}_{\mathcal{S}}{\bf T}\hspace{1cm}\equiv\textrm{P}_{\mathcal{PS}}{\bf T}\hspace{1cm}\textrm{and}\hspace{1cm}{\bf T}=\textrm{P}_{\mathcal{P}}\textrm{P}_{\mathcal{A}}{\bf T}\equiv\textrm{P}_{\mathcal{PA}}{\bf T}
	\end{equation}
	in the symmetric and anti-symmetric case, respectively. The matrices $\textrm{P}_{\mathcal{PS}}$ and $\textrm{P}_{\mathcal{PA}}$ contain thus all the information about the final form of the (symmetric in the first case and anti-symmetric in the second case) tensor: $rank(\textrm{P}_{\mathcal{PS}})$ and $rank(\textrm{P}_{\mathcal{PA}})$ give the number of independent components and each non-zero row of the matrices gives a set of constrains between the non-zero components of the tensor.
	
	If the Jahn symbol contains more that one pair of symmetric and/or anti-symmetric components, for instance [V2][V2], [V2]V2\{V2\}, \{V2\}V\{V2\}, etc..., for every pair it is possible to construct the $\mathcal{S}$ or $\mathcal{A}$ matrices and, considering the groups of index 2 formed by every such matrices and the identity, to calculate the further symmetry restrictions using exactly the same algorithm. Therefore, the final set of independent tensor components and the relations between the non-zero components can be determined by a single final matrix,
	\begin{equation}
		\label{eq:finalmat}
		\textrm{P}_{P\mathcal{S}_1\ldots \mathcal{S}_k\mathcal{A}_1\ldots \mathcal{A}_{\ell}}=\textrm{P}_{\mathcal{P}}\textrm{P}_{\mathcal{S}_1}\ldots\textrm{P}_{\mathcal{S}_k}\textrm{P}_{\mathcal{A}_1}\ldots\textrm{P}_{\mathcal{A}_{\ell}},
	\end{equation}
	where $k$ is the number of symmetric pairs of indices and $\ell$ is the number of anti-symmetric pairs.
	
	The $\mathcal{S}$ matrix defined by equations (\ref{eq:symantisym}) and (\ref{eq:SandAmatrices}) represents the interchange of two components of a tensor. However, there exist physical properties whose tensor is invariant under the permutation of 3 of more components. For instance, the \emph{optical rectification} in non-dissipative media without dispersion is described by a 3-rank tensor $\chi(0;\omega,-\omega)_{ijk}$ symmetric under the permutation of the three components (Jahn symbol [V3]) or the \emph{electric-field induced second-harmonic generation} in non-dissipative media without dispersion is described by a 4-rank tensor $\chi(2\omega;0,\omega,\omega)_{ijk\ell}$ symmetric under the permutation of the 4 indices (Jahn symbol [V4]). In general, one can consider a Jahn symbol that contains as part of the whole symbol [V$n$] with $n>2$, being $n!$ the number of possible permutations of indices. These permutations form a group isomorphic to the group of permutations of grade $n$. Let's consider that the Jahn symbol of such a tensor is V$p$[V$n$]V$(r-p-n)$, with $p\ge0$, $n>2$ integer numbers and $r$ the rank of the tensor. The $n!$ permutations can be obtained by a set of $(n-1)$ generators: the operations that represent the interchange of the components $(p+1,p+2)$, $(p+2,p+3)$,\ldots and $(p+n-1,p+n)$, constructed as shown by equations (\ref{eq:symantisym}) and (\ref{eq:SandAmatrices}). The successive multiplication of these $(n-1)$ matrices by each other gives as a result the $n!$  matrices that represent the different permutations. As these matrices form a finite group, it is possible to construct a projector equivalent to the projector of equation (\ref{eq:projector}), that finally defines an extra matrix $\textrm{P}_{\mathcal{S}n}$. This matrix can be added to the final list of matrices (\ref{eq:finalmat}) as an extra multiplicative factor.
	
Finally, it is necessary to consider those tensors whose intrinsic symmetry contains the invariance under the permutation of pairs of indices. For instance, the \emph{elastic compliance} and the \emph{elastic stiffness} tensors of rank 4 have as Jahn symbol [[V2][V2]]. These tensors are symmetric under the interchange of the first two indices, under the interchange of the last two indices (these symmetries have been considered before and they contribute to the final matrix (\ref{eq:finalmat}) with two factors) and also under the interchange of the two pairs of indices. This operation can also be represented by a matrix of the form given by equation (\ref{eq:symantisym}), but instead of the matrix $\mathcal{S}$ of dimension $9\times9$ given in equation (\ref{eq:SandAmatrices}), a $81\times81$ matrix will substitute the central $E\otimes S\otimes E$ subset in equation (\ref{eq:symantisym}). The construction of this matrix is straightforward and, together with the identity matrix, it forms a group of index 2. Following the same procedure as with the other symmetries, it will add an extra (multiplicative) term in the final matrix (\ref{eq:finalmat}).
\subsubsection{Some considerations on the algorithm}
\label{sec_extra-considerations}
To finish this section several considerations about the used algorithm should be added.

\begin{itemize} 
	\item The procedure outlined in the previous two sections allows one to calculate the symmetry reduction of any tensor that may contain intrinsic symmetries or not under any MPG. The problem reduces to the calculation of the $3^r\times3^r$ matrix of equation (\ref{eq:finalmat}). Every non-zero row of this matrix contains a relation between the tensor coefficients $T_u$ with $u=1,\ldots,3^r$ ordered according to the 1:1 map given by equation (\ref{eq:ordercomponents}). However, in general, these non-zero rows can involve intricate relations between the coefficients. To simplify the final expression of the tensor, STENSOR performs the \emph{row reduction} (or Gauss decomposition) to convert the matrix (\ref{eq:finalmat}) into its \emph{reduced row echelon form} whose main properties are: it is an upper triangular matrix, only the first $r_P=rank(\textrm{P}_{P\mathcal{S}_1\ldots \mathcal{S}_k\mathcal{A}_1\ldots \mathcal{A}_{\ell}})$ rows have at least one nonzero component, the first nonzero element in these non-zero rows is 1 and this is the only nonzero component in its column. 
	
	Using the final echelon form of the matrix (\ref{eq:finalmat}), the table of components of the tensor can be easily constructed. Each non-zero row corresponds to an independent component of the final tensor. If we denote as $\mathcal{P}^e$
	the $r_P\times3^r$-dimensional matrix that contains the non-zero rows of the echelon matrix obtained by Gauss decomposition of the matrix (\ref{eq:finalmat}), the tensor components are,
	\begin{equation}
		\label{eq:outputMPG}
		(T_1,T_2,\ldots,T_{3^r})=\left(a_1,a_2,\ldots,a_{r_P}\right)\mathcal{P}^e,
	\end{equation}
	where $a_i$ are the $r_P$ independent parameters. To assign the label to each independent coefficient, we identify the position $u$ (the column index) of the first non-zero coefficient of the row that, by construction, takes the value 1, and identify though equation (\ref{eq:ordercomponents}) the values of the indices $i_1,\ldots,i_r$. The label of the corresponding independent coefficient in STENSOR is 
	\begin{equation}
		\label{eq:outputMPG2}
		a_j=c_{i_1,\ldots,i_r}^j\hspace{1cm} j=1,\ldots,r_P.
	\end{equation}
	
	\item When the MPG belongs to the trigonal or hexagonal crystal system, usually the symmetry operations are introduced in the standard hexagonal setting, with $a=b$, $\alpha=\beta=90^{\circ}$ and $\gamma=120^{\circ}$. As the final form of the tensor components are given in an orthogonal setting, the symmetry operations and the direction ${\bf n}$ in collinear and coplanar cases are previously transformed into an orthogonal setting with $a'=b'$ and $\alpha'=\beta'=\gamma'=90^{\circ}$. In principle, there are 6 possible symmetry equivalent orientations of the final setting with respect to the original hexagonal setting. In most cases, the final form of the non-zero but dependent tensor coefficients can be expressed as linear combinations of the independent coefficients using rational coefficients. However, after a rotation of the symmetry axes by an angle multiple of $2\pi/6$, although the new directions are equivalent to the original ones, the mentioned linear combinations will include non-fractional coefficients. In the trigonal and hexagonal groups STENSOR checks the 6 symmetry equivalent pairs of orthogonal axes perpendicular to the 3-fold axis to choose the basis in which the final tensor takes the simplest form.
	
	\item The memory and cpu-time requirements in the calculation depends strongly on the rank of the given tensor. For rank-8 tensors, for instance, the matrices involved in the calculation given by equations (\ref{eq:Kronecker}) or (\ref{eq:symantisym}) are of dimension $6561\times6561$. If the order of the MPG is high, the full calculation can take a long time, and the program will stop if it does not reach the final result in a predefined time limit.
	
	 However, the memory and time requirements can be strongly reduced using \emph{sparse arrays}, implemented in most program codes of linear algebra. It is particularly useful when the matrices involved in the calculations have a relatively large number of null components. Using sparse arrays, a matrix of $n\times n$ components is described by an array that contains only the non-zero components and thus, operations as sums or multiplications of matrices are most effectively performed using these arrays, avoiding a large number of multiplications by 0.
	The use of sparse arrays is particularly useful when the matrices involved represent crystallographic symmetry operations. Except in trigonal and hexagonal point groups, when described in the standard setting (or in most reasonable settings), only 3 components of the $R$ matrices out of 9  (1/3 factor) are different from 0. In the Kronecker product of $r$ such matrices (equations (\ref{eq:Kronecker}) and (\ref{eq:symantisym})), the proportion of non-zero coefficients is $(1/3)^r$, i.e., only 
	0.01\% of coefficients for $r=8$, for instance. The matrices of the point groups in the trigonal and hexagonal crystal systems expressed in an orthogonal reference system have 3 or 5 non-zero components. The use of sparse arrays is not so advantageous in these crystal systems but still very significant. STENSOR uses sparse arrays in all the calculations with matrices.
	
	\item Except when the rank of the tensor is 2, STENSOR uses the Voigt form when the Jahn symbol contains one or several symmetric terms of type [V2]. In those cases, it uses the usual assignation $11\to1$, $22\to2$, $33\to3$,  $23,32\to4$, $13,31\to5$ and $12,21\to6$ to reduce the number of dependent components. It also uses the Voigt form of as many pairs of indices as possible for tensors that contain terms of the type [V$n$] with $n>3$. The assignation is direct, i.e., $c_{ij}\to c_{\alpha}$ with $i,j=1,2,3$ and $\alpha=1,\ldots,6$ with no assignation of factors 1/2 or 1/4 to specific tensors as it is often the case \cite{Nye1985}.
\end{itemize} \subsection{Tensor reduction under the spin point group}
\label{sec:spgreduction}
The calculation of the tensor reduction under the whole spin group can be divided into three steps, being the first two almost identical to the ones followed in the previous section to calculate the reduction under the MPG. First, the constrains imposed by \Pnte\, are expressed through a matrix equivalent to equation (\ref{eq:pointreduction}), next the constraints due to the (anti)symmetrization under the interchange of sets of indices are added as explained in section (\ref{sec:intrinsic}) and finally, as a third and final step, the additional reduction imposed by the trivial point group in collinear and coplanar spin arrangements is also added.
\subsubsection{Tensor reduction by the operations of the SpPG}
\label{sec:reduction-spg}
The reduction of the components of a tensor under the SpPG is carried out following the procedure explained in section (\ref{sec:reductionMPG}), but some expressions must be slightly modified. The constrains imposed by an operation $\{U||R\}$ of the SpPG on a rank-1 tensor depends on the type on tensor. In \citet{etxebarria2025} four different types of ferroic tensors of rank-1 tensors were described: V (polar vector, odd under inversion and even under time-reversal), $e$V (axial vector, even under both inversion and time-reversal), M (axial magnetic vector, even under inversion and odd under time-reversal) and T (toroidic moment, odd under both inversion and time-reversal). In general, the components of a tensor of rank $r$ transform as the product of rank-1 tensors, where each component can be of V- or M-type. In this section we are not considering tensors that include components that transform as T (these tensors will be analyzed in section (\ref{sec:toroidic})).

 As in section (\ref{sec:reductionMPG}), in this first step the (anti)symmetrization of the tensor under the interchange of indices is not considered and therefore, in general, the Jahn symbol to be considered is J, $e$J, $a$J or $ae$J, where J represents a set of V and M components. The procedure developed in this section is exactly the same for the four types of tensors. The V components transform under $R$ and the M components transform under $U$ when the symmetry operation of the point group $\{U||R\}$ is considered. The symmetry constrains imposed by the operation $\{U^i||R^i\}$ is thus,
 	\begin{equation}
 	\label{eq:SPGreduction}
 	T_{i_1,i_2,\ldots,i_r}=T_{i_1',i_2',\ldots,i_r'}=f^i(U|R)_{i_1'i_1}^i(U|R)_{i_2'i_2}^i\ldots (U|R)_{i_r'i_r}^iT_{i_1,i_2,\ldots,i_r},
 \end{equation}
 where $(U|R)_{i_j'i_j}^i=R_{i_j'i_j}$ or $(U|R)_{i_j'i_j}^i=U_{i_j'i_j}$ if the $j$-th term in the sequence of letters in the Jahn symbol is V or M, respectively. The expression (\ref{eq:Kronecker}) must ge generalized to,
\begin{equation}
	\label{eq:KroneckerSPG}
	\mathcal{R}^i=f^i\overbrace{(U|R)^i\otimes (U|R)^i\otimes\ldots\otimes (U|R)^i}^r.
\end{equation}
Note that, in principle, the factor $f^i$ is different in equations (\ref{eq:Kronecker}) and (\ref{eq:KroneckerSPG}). The sum of the matrices (\ref{eq:KroneckerSPG}) for all the symmetry operations of the SpPG (equation (\ref{eq:pointgroupreduction})) contains all the information about the symmetry reduction of the tensor due to the SpPG. In general, the order of the SpPG in equation (\ref{eq:KroneckerSPG}) is higher that the number of operations of the MPG in equation (\ref{eq:pointreduction}).

The (anti)symmetrization of the tensor in the context of MPG and SpPG is exactly the same: the substitution M$\to ae$V does not change the (anti)symmetric character of the tensor under the interchange of indices: both [Vn] and [Mn] transform into [Vn] (in the second case accompanied by $ae$ if $n$ is odd) and both \{V2\} and \{M2\} transform into \{V2\}. Therefore, all the factors in the final matrix given by equation (\ref{eq:finalmat}) are exactly the same, except the first one, as explained in this section.

\subsubsection{Tensor reduction by the operations of the trivial group}
\label{sec:reduction-trivial-group}
To obtain the final form of the tensor, the additional restrictions imposed by the trivial group must also be considered. In non-coplanar groups the trivial group is the identity and it does not introduce new restrictions. The matrix (\ref{eq:finalmat}) contains all the information about the tensor, which takes the form given by equations (\ref{eq:outputMPG}) and (\ref{eq:outputMPG2}). 

The restrictions added by the coplanar or collinear groups can be analyzed as follows.

\begin{itemize} 
\item In coplanar groups, apart from the identity, the only operation of the trivial group is a mirror plane perpendicular to the direction ${\bf n}=(n_1,n_2,n_3)$. As stated above, if the input parameters are not given in an orthogonal basis (usually it is the case in trigonal and hexagonal crystal systems) they are transformed into an orthogonal one, so at this point $(n_1,n_2,n_3)$ are the components of the direction in an orthogonal basis. The matrix that represents a mirror plane perpendicular to ${\bf n}$ is,
		\begin{equation}
	\label{mirrorcoplanar}
	m_{\perp\bf n}=\frac{1}{n^2}\left(
	\begin{array}{ccc}
		n^2-2n_1^2 & -2 n_1n_2 & -2 n_1n_3 \\
		-2 n_1 n_2 & n^2-2n_2^2 & -2 n_2n_3 \\
		-2 n_1 n_3 & -2 n_2n_3 & n^2-2n_3^2
	\end{array}
	\right).
\end{equation}
It is easy to check that the matrix (\ref{mirrorcoplanar}) represents the mirror plane because $m_{\perp\bf n}^2=m_{\perp\bf n}.m_{\perp\bf n}^T=1$, $\det(m_{\perp\bf n})=-1$ and $m_{\perp\bf n}{\bf n}=-{\bf n}$. To determine the restrictions imposed by the trivial group one can follow the procedure 
explained in section (\ref{sec:reductionMPG}). For this operation, in the Kronecker product (\ref{eq:KroneckerSPG}) all the terms of the product are $(U|R)\to\mathds{I}_{3}$ or $(U|R)\to U=m_{\bf n}$ in equation (\ref{mirrorcoplanar}), depending on the corresponding letter V or M, respectively, in the Jahn symbol. Together with the identity, the resulting matrix $\mathcal{R}_{m_{\perp\bf n}}$ forms a group of order 2 and the corresponding projector (\ref{eq:projector}) can be defined. Finally, the symmetry restrictions of the trivial group are included in the matrix,
\begin{equation}
	P_{m_{\perp\bf n}}=\frac{1}{2}\left(\mathds{I}_{3^r}+\mathcal{R}_{m_{\perp\bf n}}\right),
\end{equation}
which can be added as a multiplicative factor to the matrix (\ref{eq:finalmat}) that contains all the symmetry restrictions considered so far. After the row reduction of the final matrix, the final the form of the output is given by equations (\ref{eq:outputMPG}) and (\ref{eq:outputMPG2}).

\item In the collinear case the trivial group contains a rotation axis of order infinite and a set of infinite parallel planes that contain the axis. The trivial group can be expressed as the direct product of two subgroups: the subgroup that contains all the operations that represent rotations around the unique axis (of arbitrary angle) and a subgroup of order 2 that contains the identity and a mirror plane
\begin{equation}
	\label{eq:descomptrivial}
	\infty_{\bf n}m=(E+m_{\parallel\bf n})\otimes\infty_{\bf n},
\end{equation} 
where $m_{\parallel\bf n}$ is any mirror plane that satisfies $m_{\parallel\bf n}{\bf n}={\bf n}$. The symmetry reduction of the tensor due to the trivial group has thus two contributions that can be analyzed separately: on the one hand the restrictions under the subgroup formed by the identity and a mirror plane and on the other hand the restrictions imposed by the rotations $\infty_{\bf n}$.

The algorithm to calculate the restrictions due to the mirror plane is exactly the same as the one used in the coplanar case. One matrix that represents a mirror plane that keeps invariant the vector ${\bf n}=(n_1,n_2,n_3)$ is,
	\begin{equation}
		\label{eq:matrixmparallel}
		m_{\parallel\bf n}=\left\{\begin{array}{cl}\frac{1}{n_1^2+n_2^2}\left(\begin{array}{ccc}
			n_1^2-n_2^2&2n_1n_2&0\\
			2n_1n_2&n_2^2-n_1^2&0\\
			0&0&n_1^2+n_2^2
		\end{array}\right)&\textrm{if }n_1^2+n_2^2\ne0\\
		\left(\begin{array}{ccc}
			-1&0&0\\
			0&1&0\\
			0&0&1
		\end{array}\right)&\textrm{if }n_1^2+n_2^2=0
		\end{array}\right..
\end{equation}
After the calculation of the $\mathcal{R}_{m_{\parallel\bf n}}$ matrix as the Kronecker product (\ref{eq:KroneckerSPG}) with $(U|R)\to\mathds{I}_{3}$ or $(U|R)\to U=m_{\parallel\bf n}$, depending on the corresponding letter V or M, respectively, in the Jahn symbol, the matrix
\begin{equation}
	\label{eq:mparallel}
	P_{m_{\parallel\bf n}}=\frac{1}{2}\left(\mathds{I}_{3^r}+\mathcal{R}_{m_{\parallel\bf n}}\right)
\end{equation}
is added as a multiplicative factor to the matrix (\ref{eq:finalmat}).

To determine the restrictions imposed by the second group in the direct product (\ref{eq:descomptrivial}) the strategy used by STENSOR is different from the one followed with all other symmetries in sections (\ref{sec:mpgreduction}) and (\ref{sec:spgreduction}). The order of the group $\infty_{\bf n}$ is infinite and it is not possible to define a projector as in equation (\ref{eq:projector}). For this group, STENSOR calculates first the final form of the tensor due to the rest of symmetry operations: the operations of \Pnte\,, the (anti)symmetric sets (if any) and the mirror plane (\ref{eq:matrixmparallel}). The partially reduced form of the tensor takes the form of equations 
(\ref{eq:outputMPG}) and (\ref{eq:outputMPG2}). Then, on this final form, STENSOR establishes the set of linear equations (\ref{eq:linearequations}) for the $\mathcal{R}$ matrix of one generator (arbitrary rotation of angle $\varphi$). In general, these linear equations will introduce relations between the so far independent $c_{i_1,\ldots,i_r}^j$ parameters. Finally STENSOR reconstructs the output taking into consideration these relations.

The matrix that represents a rotation of angle $\varphi$ around an axis parallel to ${\bf n}=(n_1,n_2,n_3)$ is,
\begin{footnotesize}
\begin{equation}
	\label{eq:rotation}
C_{\varphi}=\frac{1}{n^2}\left(
\begin{array}{ccc}
n_1^2+\left(n_2^2+n_3^2\right)\cos\varphi &  n_1n_2\left(1-\cos\varphi\right)-nn_3\sin\varphi & n_1n_3\left(1-\cos\varphi\right)+nn_2\sin\varphi \\
n_1n_2\left(1-\cos\varphi\right)+nn_3\sin\varphi & n_2^2+\left(n_1^2+n_3^2\right)\cos\varphi &  n_2n_3\left(1-\cos\varphi\right)-nn_1\sin\varphi \\
n_1n_3\left(1-\cos\varphi\right)-nn_2\sin\varphi & n_2n_3\left(1-\cos\varphi\right)+nn_1\sin\varphi & n_3^2+\left(n_1^2+n_2^2\right)\cos\varphi
\end{array}
\right).
\end{equation}
\end{footnotesize}
It can be checked that it is the right matrix: $C_{\varphi}.C_{\varphi}^T=1$, $\det(C_{\varphi})=1$, $C_{\varphi}.C_{\varphi}=C_{2\varphi}$ and $C_{\varphi}{\bf n}={\bf n}$.

As a generator of the group, it can be considered an infinitesimal angle $\varphi\ll1$ and perform a series expansion of the matrix (\ref{eq:rotation}) keeping only up to the linear term. The matrix is
\begin{equation}
		C_{\varphi}\simeq\mathds{I}_3+\varphi \omega,
\end{equation}
with
\begin{equation}
	\label{eq:omega}
	\omega=\frac{1}{n}\left(
	\begin{array}{ccc}
		0 &  -n_3 & n_2 \\
		n_3 & 0 &  -n_1 \\
		-n_2 & n_1 & 0
	\end{array}
	\right).
\end{equation}
The Kronecker product (\ref{eq:KroneckerSPG}) up to the linear term takes the form
\begin{equation}
	\mathcal{R}^{\infty}=\mathds{I}_{3^r}+\varphi\Omega,
\end{equation}
being $\Omega$ a $3^r\times3^r$-dimensional numerical (independent of $\varphi$) matrix.
Using vector notation, the linear set of equations (\ref{eq:linearequations}) takes de form,
\begin{equation}
	\label{eq:homogeneous}
	{\bf T}=\left(\mathds{I}_{3^r}+\varphi\Omega\right){\bf T}\hspace{1cm}\rightarrow\hspace{1cm}\Omega{\bf T}=0.
\end{equation}
Taking the last form (\ref{eq:outputMPG}) of ${\bf T}$, the set of homogeneous linear equations (\ref{eq:homogeneous}) can be written as,
\begin{equation}
	\label{eq:homOmega}
	\left(a_1,a_2,\ldots,a_r\right)\mathcal{P}^e\Omega^T=0.
\end{equation}
In general, this set of equations will introduce linear relations between the \emph{independent} coefficients $a_i$. Using these relations between the coefficients, the final form of the tensor is,
\begin{equation}
	\label{eq:finalreduction}
	(T_1,T_2,\ldots,T_{3^r})=\left(a'_1,a'_2,\ldots,a'_{r_P'}\right)\mathcal{P'}^e,
\end{equation}
with $r_P'\le r_P$ and being $\mathcal{P'}^e$ an echelon matrix. As in equations (\ref{eq:outputMPG}) and (\ref{eq:outputMPG2}), the label of the $a'_j$ coefficients shown by STENSOR is $a'_j=c_{i_1,\ldots,i_r}^j$, being $i_1,\ldots,i_r$ the set of indices that correspond to the first non-zero value of the $j^{\textrm{th}}$ row of $\mathcal{P'}^e$.
\end{itemize}
\subsubsection{Symmetry constrains of tensors that contain toroidic components}
\label{sec:toroidic}
The moment of magnetization or \emph{toroidic} moment ${\bf T}={\bf r}\times{\bf M}$ is odd both under spatial inversion and under time-reversal symmetry. The transformation properties of these quantities involve both $R$ space and $U$ spin operations and can be obtained as the antisymmetric part of the magnetoelectric tensor (direct or inverse effect) \cite{Spaldin2008}. Following \citet{etxebarria2025} and taking the inverse magnetoelectric effect as the reference tensor, whose Jahn symbol is MV, we denote the Jahn symbol of the toroidic moment as \{MV\}, reflecting the fact that it transforms as the antisymmetric part of the magnetoelectric tensor $\alpha_{ij}^T$. STENSOR performs the symmetry reduction of a tensor that contains, as a part of its Jahn symbol, one of several components that transform as the toroidic moment, in two steps:
\begin{itemize}
	\item First, it determines the symmetry reduction of the tensor $T^{\prime}$ with the Jahn symbol that results after the substitution \{MV\}$\to$MV, following the procedure described in section \ref{sec:mpgreduction} for the MPG and section \ref{sec:spgreduction} for the SpPG. The output contains the corresponding two tables.
	\item Taking the general form for the SpPG (second table in the previous step), STENSOR makes the antisymmetric reduction using the Levi-Civita tensor \cite{etxebarria2025},
	\begin{equation}
		\label{eq:levicivita}
		T_{i_1,\ldots, i_{j-1},k,i_{j+2},\ldots,i_r}=\frac{1}{2}\varepsilon_{k,i_j,i_{j+1}}T_{i_1,\ldots, i_{j-1},i_j,i_{j+1},i_{j+2},\ldots,i_r}^{\prime},
	\end{equation}
	where it has been assumed that the M and V symbols inside \{MV\} occupy the positions $i_j$ and $i_{j+1}$ in the Jahn symbol. If the Jahn symbol contains more than one \{MV\} term, it must be performed a reduction of the type given by equation (\ref{eq:levicivita}) for each term.
\end{itemize}
At the end of the output, a third table shows the general form of the quantity that contains toroidic terms. \section{Example: symmetric spin contribution to the Hall effect in NiCr$_2$O$_4$}
\label{sec:example}
This section includes a detailed application of the algorithm explained in section \ref{sec:methods} using an example. The physical property chosen is the symmetric part of the Hall effect (or linear magnetorresistence) $R_{ijk}^s$, whose tensor has Jahn symbol [V2]M and the compound is  NiCr$_2$O$_4$ \cite{Tomiyasu2004} (entry 0.4 in MAGNDATA). The identified spin space group \cite{Chen2024} is collinear with symbol $P\,^{1}4_1/\,^{1}a\,^{1}m\,^{1}d\,^{\infty_{110}m}1$ (N. 141.141.1.1). The non-trivial group \Pnt\, is $^{1}4/\,^{1}m\,^{1}m\,^{1}m$ and the spin-only group is $^{\infty_{110}m}1$ with the spins aligned along the $(1,1,0)$ direction.

Fig. \ref{fig:menu} in the main text shows the input page of STENSOR where the data of the example have been introduced. The red text and the arrows have been added to the figure to show the data introduced by the user. As it can be seen in the figure, the following three operations have been introduced as generators of  \Pnt\,
\begin{equation}
	\begin{array}{cc}
		-y,x, z,+1,u,v,w&\{1||4_{001}^{+}\}\\
		 x,y,-z,+1,u,v,w&\{1||m_{001}\}\\
		-x,y, z,+1,u,v,w&\{1||m_{100}\}
	\end{array}
\end{equation}
It has been assumed that the basis vectors in the orbital and spin spaces are exactly the same. If the $U$ and $R$ matrices of the spin and orbital spaces, respectively, were expressed in different bases, clicking on the button \emph{spin basis} a $3\times3$ table would emerge. In that table the user can introduce the components of the matrix that relates both reference systems. By default the program assumes that this matrix is the identity.

\subsection{Determination of the MPG}
The complete set of operations of \Pnt\, calculated by STENSOR are,
\begin{equation}\label{eq:operations}\begin{array}{llllllll}
		\{1||1\}&\{1||4_{001}^{+}\}&\{1||2_{001}\}&\{1||4_{001}^{-}\}&\{1||\overline{1}\}&\{1||\overline{4}_{001}^{+}\}&\{1||m_{001}\}&\{1||\overline{4}_{001}^{-}\}\\
		\{1||m_{100}\}&\{1||m_{1\overline{1}0}\}&\{1||m_{010}\}&\{1||m_{110}\}&\{1||2_{100}\}&\{1||2_{1\overline{1}0}\}&\{1||2_{010}\}&\{1||2_{110}\}
\end{array}\end{equation}
Note that the complete set of operations are the direct product of the operations in the list (\ref{eq:operations}) and the operations $\{U||1\}\}$ of the trivial group.

\begin{figure}[ht]
	\centering
	\includegraphics[width=0.8\textwidth]{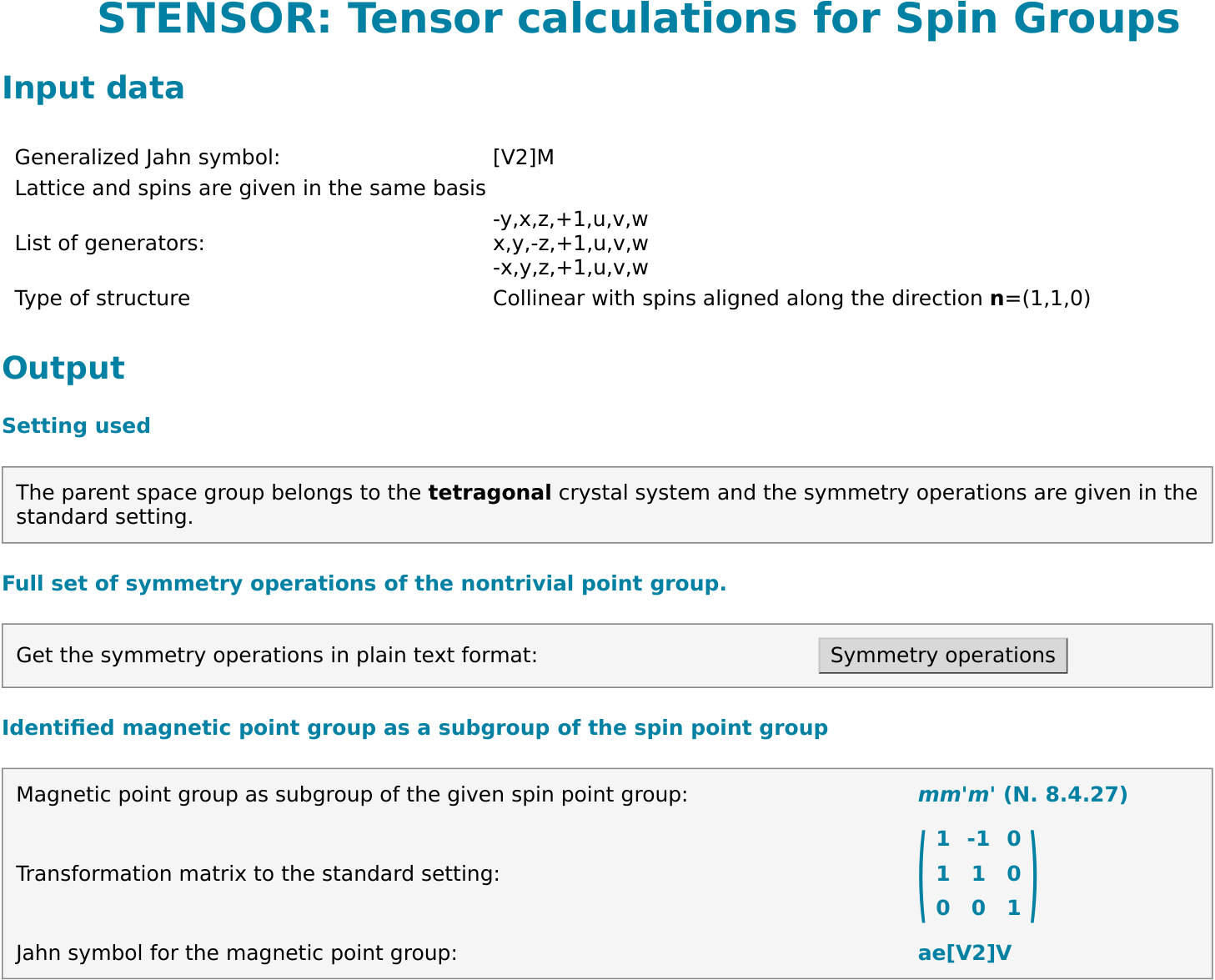}
	\caption{\label{fig:exampleoutputMPG}First part of STENSOR's output for the input given in Fig. (\ref{fig:menu}). It shows the input data, a link to the complete list of symmetry operations of \Pnt\, in plain text, information about the crystal system, the identified MPG, the transformation matrix from the setting given by the user to the standard setting of the MPG and the Jahn symbol for the MPG.}
\end{figure}

Following the procedure developed in section (\ref{sec:methods}) the program selects from the list (\ref{eq:operations}) those that fulfill the condition (\ref{eq:condcollinear}) with ${\bf n}=(1,1,0)$ in this particular example. As the spin operation $U$ is the identity in all the cases, the condition transforms into $R{\bf n}=\theta{\bf n}$ with $\theta=\pm1$ for all the operations. Only the following 8 operations fulfill one of the conditions,
\begin{footnotesize}
\begin{equation}\label{eq:MPG}\begin{array}{l|l|l|l}
		1{\bf n}={\bf n}\rightarrow\{1,1\}&2_{001}{\bf n}=-{\bf n}\rightarrow\{2_{001},-1\}&\overline{1}{\bf n}=-{\bf n}\rightarrow\{\overline{1},1\}&m_{001}{\bf n}={\bf n}\rightarrow\{m_{001},-1\}\\
		m_{1\overline{1}0}{\bf n}={\bf n}\rightarrow\{m_{1\overline{1}0},-1\}&m_{110}{\bf n}=-{\bf n}\rightarrow\{m_{110},1\}&\{2_{1\overline{1}0},-1\}{\bf n}=-{\bf n}\rightarrow\{2_{1\overline{1}0},-1\}&2_{110}{\bf n}={\bf n}\rightarrow\{2_{110},1\}
\end{array}\end{equation}
\end{footnotesize}
On the right side of each $R{\bf n}=\theta{\bf n}$ condition it has been shown the resulting operation of the MPG in the usual notation $\{R,\theta\}$. The 8 operations form the MPG $mm'm'$ (N. 8.4.27), but they are not expressed in the standard setting of this MPG. The two planes parallel to the $z$ axis are rotated 45$^{\circ}$ with respect to the planes in the standard description. Together with the identification of the symbol and sequential number of the MPG, STENSOR gives also a transformation matrix to the standard setting. Fig. (\ref{fig:exampleoutputMPG}) shows the first part of the output given by STENSOR. 

The first part of the output includes the input parameters, the identification of the crystal system, information about the setting used for the orbital part of the symmetry operations and a link to the complete set of symmetry operations of \Pnt\, (list of operations in equation (\ref{eq:operations}) in the $x,y,z,+1,u,v,w$ format). Just below the information related with the input data, the program shows the identified MPG together with the transformation matrix $P$ that relates the operations in (\ref{eq:MPG}) with the operations of the standard setting of $mm'm'$, i.e., $\{R_s,\theta\}=\{P^{-1}RP,\theta\}$, where $R$ is the rotational part of the operations in equation (\ref{eq:MPG}) and $R_s$ is the corresponding operation in the standard description. The output also shows the Jahn symbol in the context of MPG of the tensor given. In this example, doing the substitution M$\to ae$V, the Jahn symbol to be used to calculate the symmetry reduction by the MPG is $ae$[V2]V.

\subsection{Tensor reduction by the MPG}
\label{sec:exMPG}
Next, the program performs the tensor reduction by the MPG that, as explained in sections (\ref{sec:reductionMPG}) and (\ref{sec:intrinsic}), consists in the determination of the matrices $\textrm{P}_{\mathcal{P}}$ in equation (\ref{eq:pointgroupreduction}) and $\textrm{P}_{\mathcal{S}}$ in equation (\ref{eq:simanti}), which reproduce the symmetry reduction due to the point group and the intrinsic symmetry of the tensor, respectively. To calculate the first one, the program first does the Kronecker products (\ref{eq:Kronecker}) of the 8 operations of the MPG in equation (\ref{eq:MPG}) with $r=3$ and $f^i=\theta_i\det(R^i)$ for the tensor chosen $ae$[V2]V. The matrix (\ref{eq:pointgroupreduction}) is just the sum of these 8 matrices. The matrix $\mathcal{S}$ that represents the interchange of the first two indices of the tensor (equations (\ref{eq:symantisym}) and (\ref{eq:SandAmatrices})) is in this case $\mathds=S\otimes\mathds{I}$. This allows to construct the matrix $\textrm{P}_{\mathcal{S}}$ in equation(\ref{eq:simanti}). The matrix that allows to determine the full reduction of the $ae$[V2]V for the MPG of equation (\ref{eq:MPG}) is the product of both matrices (equation (\ref{eq:simanti})). After the row reduction of this product, and keeping only the non-zero rows, the final matrix has only 5 rows, shown below the line in table (\ref{t-exampleMPG1}). The table also lists the 27 coefficients of the tensor above the line, following the convention given by equation (\ref{eq:ordercomponents}).

\begin{footnotesize}
\begin{table}[h]
	\caption{\label{t-exampleMPG1} Above the line, the components of a rank-3 tensor are indicated (the indices are aligned vertically to save horizontal space), in the order given by expression (\ref{eq:ordercomponents}). Below the line, the table contains the non-zero components of the matrix that gives the symmetry reduction of the tensor with Jahn symbol ae[V2]V under the MPG with elements in equation (\ref{eq:operations}) (after the row reduction).}
\begin{tabular}{ccccccccccccccccccccccccccc}
$c$ & $c$ & $c$ & $c$ & $c$ & $c$ & $c$ & $c$ & $c$ & $c$ & $c$ & $c$ & $c$ & $c$ & $c$ & $c$ & $c$ & $c$ & $c$ & $c$ & $c$ & $c$ & $c$ & $c$ & $c$ & $c$ & $c$\\
1  &  1  &  1  &  1  &  1  &  1  &  1  &  1  &  1  &  2  &  2  &  2  &  2  &  2  &  2  &  2  &  2  &  2  &  3  &  3  &  3  &  3  &  3  &  3  &  3  &  3  &  3\\
1  &  1  &  1  &  2  &  2  &  2  &  3  &  3  &  3  &  1  &  1  &  1  &  2  &  2  &  2  &  3  &  3  &  3  &  1  &  1  &  1  &  2  &  2  &  2  &  3  &  3  &  3\\
1  &  2  &  3  &  1  &  2  &  3  &  1  &  2  &  3  &  1  &  2  &  3  &  1  &  2  &  3  &  1  &  2  &  3  &  1  &  2  &  3  &  1  &  2  &  3  &  1  &  2  &  3\\    
\hline
1 & 0 & 0 & 0 & 0 & 0 & 0 & 0 & 0 & 0 & 0 & 0 & 0 & 1 & 0 & 0 & 0 & 0 & 0 & 0 & 0 & 0 & 0 & 0 & 0 & 0 & 0 \\
0 & 1 & 0 & 0 & 0 & 0 & 0 & 0 & 0 & 0 & 0 & 0 & 1 & 0 & 0 & 0 & 0 & 0 & 0 & 0 & 0 & 0 & 0 & 0 & 0 & 0 & 0 \\
0 & 0 & 0 & 1 & 1 & 0 & 0 & 0 & 0 & 1 & 1 & 0 & 0 & 0 & 0 & 0 & 0 & 0 & 0 & 0 & 0 & 0 & 0 & 0 & 0 & 0 & 0 \\
0 & 0 & 0 & 0 & 0 & 0 & 0 & 0 & 1 & 0 & 0 & 0 & 0 & 0 & 0 & 0 & 0 & 1 & 0 & 0 & 1 & 0 & 0 & 1 & 0 & 0 & 0 \\
0 & 0 & 0 & 0 & 0 & 0 & 0 & 0 & 0 & 0 & 0 & 0 & 0 & 0 & 0 & 0 & 0 & 0 & 0 & 0 & 0 & 0 & 0 & 0 & 1 & 1 & 0 
\end{tabular}
\end{table}
\end{footnotesize}

Using the Voigt correspondence for the two first indices, reordering the columns to align the newly defined components, and performing again the row reduction of the matrix, the final result is shown in table (\ref{t-exampleMPG2}). Every row corresponds to an independent component of the tensor. On the left of the table, it has been added another column with the label assigned to each independent coefficient. The natural election of the label is the tensor coefficient that corresponds to the first non-zero value in the row (it is always 1 by construction of the echelon form of the matrix). The final value of every element of the tensor is the sum of matrix elements of its column, multiplied by the corresponding coefficient.
\begin{footnotesize}
	\begin{table}[h]
	\caption{\label{t-exampleMPG2} Reduced form of table \ref{t-exampleMPG1}, using Voigt-like notation. The columns have been reordered to keep the 1,2,\ldots sequence of indices and the row reduction has been applied again. On the left, a label has been assigned to each independent row.}
		\begin{tabular}{c|cccccccccccccccccc}
			&  $c$ & $c$ & $c$ & $c$ & $c$ & $c$ & $c$ & $c$ & $c$ & $c$ & $c$ & $c$ & $c$ & $c$ & $c$ & $c$ & $c$ & $c$\\
			&  1  &  1  &  1  &  2  &  2  &  2  &  3  &  3  &  3  &  4  &  4  &  4  &  5  &  5  &  5  &  6  &  6  &  6\\
			&  1  &  2  &  3  &  1  &  2  &  3  &  1  &  2  &  3  &  1  &  2  &  3  &  1  &  2  &  3  &  1  &  2  &  3\\    
			\hline
			$c_{11}$  &	1 & 0 & 0 & 0 & 1 & 0 & 0 & 0 & 0 & 0 & 0 & 0 & 0 & 0 & 0 & 0 & 0 & 0\\
			$c_{12}$  &	0 & 1 & 0 & 1 & 0 & 0 & 0 & 0 & 0 & 0 & 0 & 0 & 0 & 0 & 0 & 0 & 0 & 0\\
			$c_{31}$  &	0 & 0 & 0 & 0 & 0 & 0 & 1 & 1 & 0 & 0 & 0 & 0 & 0 & 0 & 0 & 0 & 0 & 0\\
			$c_{43}$  &	0 & 0 & 0 & 0 & 0 & 0 & 0 & 0 & 0 & 0 & 0 & 1 & 0 & 0 & 1 & 0 & 0 & 0\\
			$c_{61}$  &	0 & 0 & 0 & 0 & 0 & 0 & 0 & 0 & 0 & 0 & 0 & 0 & 0 & 0 & 0 & 1 & 1 & 0
		\end{tabular}
	\end{table}
\end{footnotesize}
Fig. (\ref{fig:exampleoutputMPG}) shows the second part of the output given by STENSOR and displays the final form of the tensor (table (\ref{t-exampleMPG2})) under the reduction of the MPG and the intrinsic symmetry, expressed in the Voigt-like form.
\begin{figure}[ht]
	\centering
	\includegraphics[width=0.8\textwidth]{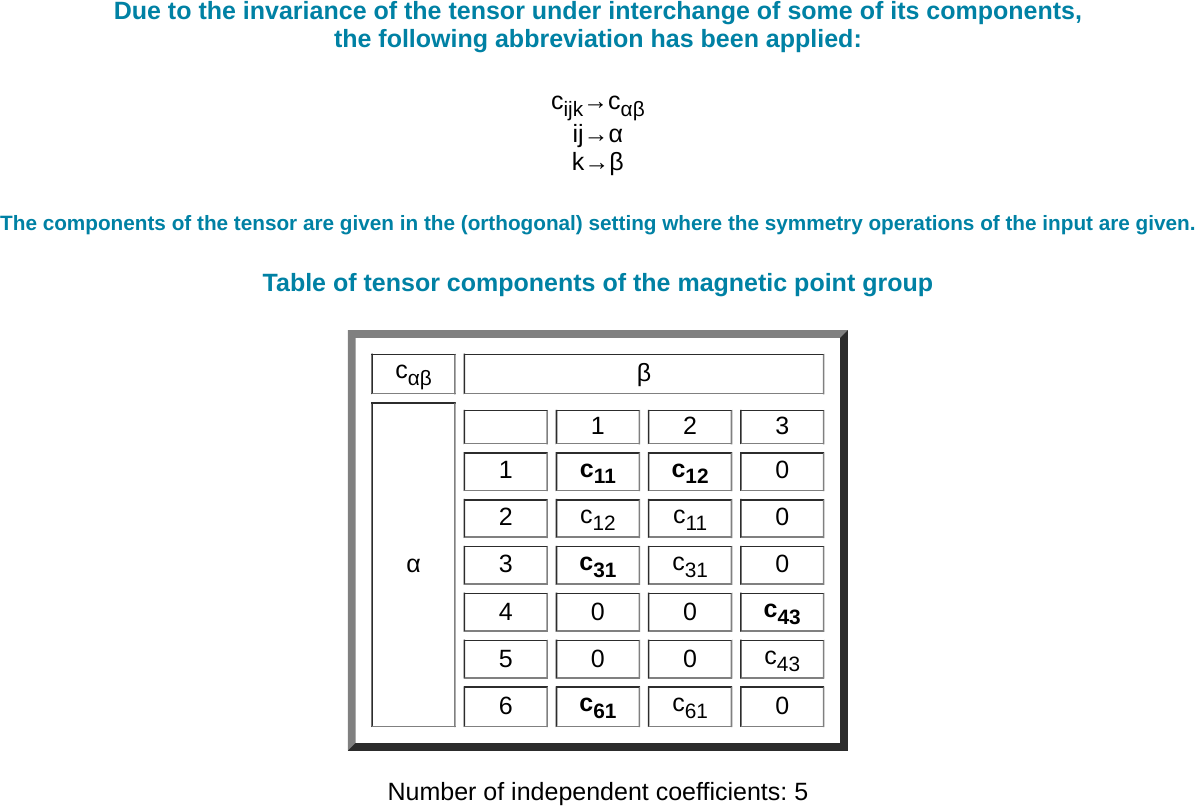}
	\caption{\label{fig:exampleoutputredMPG}Symmetry reduced Voigt-like form of the tensor with Jahn symbol ae[V2]V under the MPG formed by the operations of equation (\ref{eq:MPG}). The independent coefficients are written in bold.}
\end{figure}
\subsection{Tensor reduction by the SpPG}
Following the method described in section (\ref{sec:spgreduction}), STENSOR calculates three matrices: 
\begin{itemize}
	\item The matrix (\ref{eq:pointgroupreduction}) which contains the information about the symmetry restrictions imposed by \Pnt. The matrix $\mathcal{R}^s$ for the tensor chosen [V2]M is,
	\begin{equation}
		\mathcal{R}^s=R^s\otimes R^s\otimes U^s
	\end{equation}
	and the summation must be calculated for all the 16 members of the SpPG (\ref{eq:operations}).
	\item The matrix $\textrm{P}_{\mathcal{S}}$ in equation (\ref{eq:simanti}) which contains the information about the restrictions imposed by the intrinsic symmetry. This matrix is exactly the same as the matrix used in section \ref{sec:exMPG} to calculate the restrictions under the MPG.
	\item The matrix $P_{m_{\parallel\bf n}}$ in equation (\ref{eq:mparallel}) that gives the restrictions of a single mirror plane of the trivial group in the collinear case. In this particular case, ${\bf n}=(1,1,0)$, the matrix (\ref{eq:matrixmparallel}) reduces to,
	\begin{equation}
		m_{\parallel\bf n}=\left(\begin{array}{ccc}
			0&1&0\\
			1&0&0\\
			0&0&1
		\end{array}\right)
	\end{equation}
	and the $\mathcal{R}_{m_{\parallel\bf n}}$ matrix is,
	\begin{equation}
			\mathcal{R}_{m_{\parallel\bf n}}=\mathds{I}_3\otimes \mathds{I}_3\otimes m_{\parallel\bf n}
	\end{equation}
\end{itemize}
The matrix that gives the partial reduction of the tensor by the SpPG (before the application of the reduction imposed by the axis $\infty_{\bf n}$) after the row reduction, and keeping only the non-zero rows of the matrix, the result is shown in table (\ref{t-exampleSG1}).

\begin{footnotesize}
\begin{table}[h]
	\caption{\label{t-exampleSG1} Above the line, the components of a rank-3 tensor are indicated (the indices are aligned vertically to save horizontal space), in the order given by expression (\ref{eq:ordercomponents}). Below the line, the table contains the non-zero components of the matrix $\mathcal{P}^e$ in equation (\ref{eq:outputMPG}) that gives the partial symmetry reduction of the tensor with Jahn symbol [V2]M under the SpPG with elements in equation (\ref{eq:operations}) (after the row reduction). To get the final reduction, the additional constraints imposed by the infinite axis must be added.}
		\begin{tabular}{ccccccccccccccccccccccccccc}
			$c$ & $c$ & $c$ & $c$ & $c$ & $c$ & $c$ & $c$ & $c$ & $c$ & $c$ & $c$ & $c$ & $c$ & $c$ & $c$ & $c$ & $c$ & $c$ & $c$ & $c$ & $c$ & $c$ & $c$ & $c$ & $c$ & $c$\\
			1  &  1  &  1  &  1  &  1  &  1  &  1  &  1  &  1  &  2  &  2  &  2  &  2  &  2  &  2  &  2  &  2  &  2  &  3  &  3  &  3  &  3  &  3  &  3  &  3  &  3  &  3\\
			1  &  1  &  1  &  2  &  2  &  2  &  3  &  3  &  3  &  1  &  1  &  1  &  2  &  2  &  2  &  3  &  3  &  3  &  1  &  1  &  1  &  2  &  2  &  2  &  3  &  3  &  3\\
			1  &  2  &  3  &  1  &  2  &  3  &  1  &  2  &  3  &  1  &  2  &  3  &  1  &  2  &  3  &  1  &  2  &  3  &  1  &  2  &  3  &  1  &  2  &  3  &  1  &  2  &  3\\    
			\hline
			1 & 1 & 0 & 0 & 0 & 0 & 0 & 0 & 0 & 0 & 0 & 0 & 1 & 1 & 0 & 0 & 0 & 0 & 0 & 0 & 0 & 0 & 0 & 0 & 0 & 0 & 0 \\
			0 & 0 & 1 & 0 & 0 & 0 & 0 & 0 & 0 & 0 & 0 & 0 & 0 & 0 & 1 & 0 & 0 & 0 & 0 & 0 & 0 & 0 & 0 & 0 & 0 & 0 & 0 \\
			0 & 0 & 0 & 0 & 0 & 0 & 0 & 0 & 0 & 0 & 0 & 0 & 0 & 0 & 0 & 0 & 0 & 0 & 0 & 0 & 0 & 0 & 0 & 0 & 1 & 1 & 0 \\
			0 & 0 & 0 & 0 & 0 & 0 & 0 & 0 & 0 & 0 & 0 & 0 & 0 & 0 & 0 & 0 & 0 & 0 & 0 & 0 & 0 & 0 & 0 & 0 & 0 & 0 & 1 
		\end{tabular}
	\end{table}
\end{footnotesize}
	The provisional form of the components of the tensor are,
	\begin{equation}
		(T_1,\ldots,T_{27})=(c_{111},c_{113},c_{331},c_{333})\mathcal{P}^e
	\end{equation}
	with $\mathcal{P}^e$ being the $4\times27$ matrix shown below the line in table (\ref{t-exampleSG1}).
	
\begin{figure}[ht]
	\centering
	\includegraphics[width=0.8\textwidth]{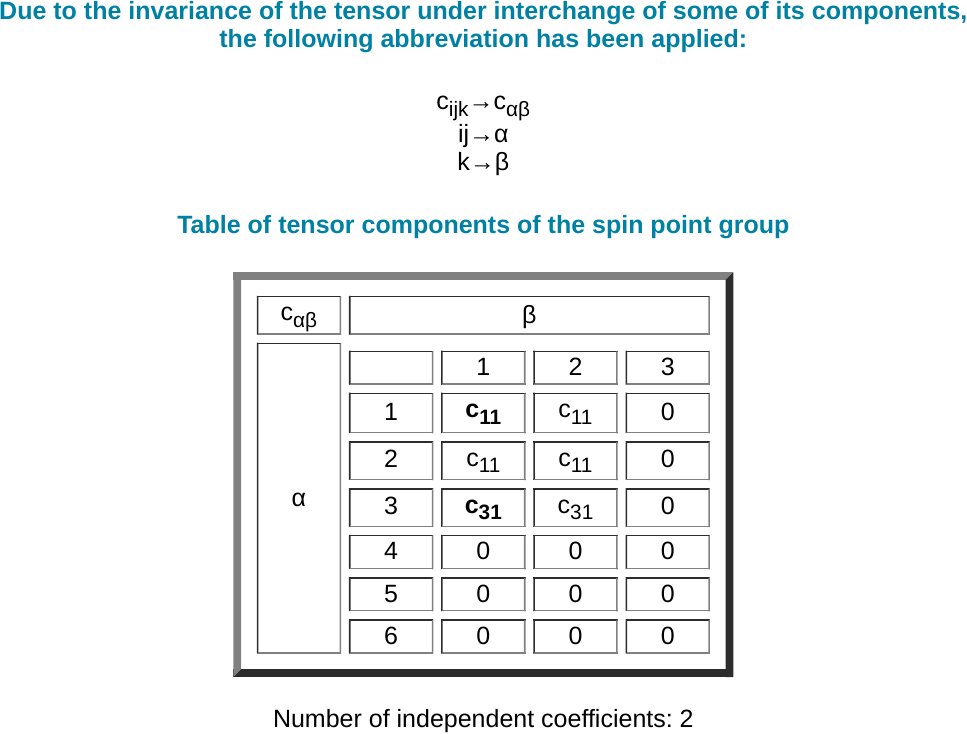}
	\caption{\label{fig:exampleoutputredSPG}Symmetry reduced Voigt-like form of the tensor with Jahn symbol [V2]M under the SpPG formed by the operations of equation (\ref{eq:operations}). The independent coefficients are written in bold.}
\end{figure}

	Finally, to get the final for of the tensor under the collinear spin group the symmetry restrictions imposed by the infinite axis must be added. The $\omega$ matrix of equation (\ref{eq:omega}) is,
	\begin{equation}
		\omega=\left(
		\begin{array}{ccc}
			0 &  0 & \frac{1}{\sqrt{2}} \\
			0 & 0 &  -\frac{1}{\sqrt{2}} \\
			-\frac{1}{\sqrt{2}} & \frac{1}{\sqrt{2}} & 0
		\end{array}
		\right)	
	\end{equation}
	and the $\Omega$ matrix necessary to write the homogeneous linear equation (\ref{eq:homOmega}) is,
	\begin{equation}
		\Omega=\frac{1}{\varphi}\left(\mathcal{R}^{\infty}-\mathds{I}_{3^r}\right)=\mathds{I}_3\otimes\mathds{I}_3\otimes\omega
	\end{equation}
	The linear set of equations (\ref{eq:homOmega}) is,
	\begin{equation}
	(c_{111},c_{113},c_{331},c_{333})\mathcal{P}^e\Omega
\end{equation}
whose general solution is $c_{113}=c_{333}=0$. The table (\ref{t-exampleSG1}) transforms thus into table (\ref{t-exampleSG2}).
\begin{footnotesize}
	\begin{table}[h]
		\caption{\label{t-exampleSG2} Above the line, the components of a rank-3 tensor are indicated (the indices are aligned vertically to save horizontal space), in the Voigt-like notation. Below the line, the table contains the non-zero components of the matrix $\mathcal{P'}^e$ in equation (\ref{eq:finalreduction}). This matrix gives the final symmetry reduction of the tensor with Jahn symbol [V2]M under the SpPG with elements in equation (\ref{eq:operations}) (after the row reduction).}
		\begin{tabular}{cccccccccccccccccc}
			$c$ & $c$ & $c$ & $c$ & $c$ & $c$ & $c$ & $c$ & $c$ & $c$ & $c$ & $c$ & $c$ & $c$ & $c$ & $c$ & $c$ & $c$\\
			1  &  1  &  1  &  2  &  2  &  2  &  3  &  3  &  3  &  4  &  4  &  4  &  5  &  5  &  5  &  6  &  6  &  6\\
			1  &  2  &  3  &  1  &  2  &  3  &  1  &  2  &  3  &  1  &  2  &  3  &  1  &  2  &  3  &  1  &  2  &  3\\    
			\hline
			1 & 1 & 0 & 1 & 1 & 0 & 0 & 0 & 0 & 0 & 0 & 0 & 0 & 0 & 0 & 0 & 0 & 0 \\
			0 & 0 & 0 & 0 & 0 & 0 & 1 & 1 & 0 & 0 & 0 & 0 & 0 & 0 & 0 & 0 & 0 & 0 \\
		\end{tabular}
	\end{table}
\end{footnotesize}
The final tensor has only 2 independent parameters, chosen as $c_{11}$ and $c_{31}$ in the Voigt form, and 6 non-zero components, $c_{11}=c_{12}=c_{21}=c_{22}$ and $c_{31}=c_{32}$. Fig. (\ref{fig:exampleoutputredSPG}) reproduces STENSOR's third part of the output.

\clearpage
\bibliography{references} \end{document}